\documentclass[twocolumn]{aastex631}
\usepackage{wrapfig,lipsum,booktabs}
\usepackage{verbatim}
\usepackage{natbib}
\usepackage{silence}
\usepackage{longtable}
\usepackage{rotating}
\usepackage{amsmath}
\usepackage[flushleft]{threeparttable}
\usepackage[graphicx]{realboxes}
\usepackage{array}
\usepackage{graphicx}
\usepackage{float}

\shorttitle{Atmospheres of Early-type Galaxies}
\shortauthors{Frisbie et al.}

\begin{document}

\title{Relationships Between Stellar Velocity Dispersion and the Atmospheres of Early-Type Galaxies}

\author{Rachel L.S. Frisbie}
\author{Megan Donahue}
\author{G. Mark Voit}
\affil{Physics and Astronomy Department, Michigan State University, East Lansing, MI 48824-2320, USA}
\author{Kiran Lakhchaura}
\affil{MTA-ELTE Astrophysics Research Group, P\'{a}zm\'{a}ny P\'{e}ter s\'{e}t\'{a}ny 1/A, Budapest, 1117, Hungary}
\author{Norbert Werner}
\affil{Department of Theoretical Physics and Astrophysics, Faculty of Science, Masaryk University, Kotl\'{a}\v{r}sk\'{a} 2, Brno, 611 37, Czech Republic}
\author{Ming Sun}
\affil{ Department of Physics and Astronomy, University of Alabama in Huntsville, Huntsville, AL 35899, USA}

\begin{abstract}
The \citet{Voit2019} black hole feedback valve model predicts relationships between stellar velocity dispersion and atmospheric structure among massive early-type galaxies. In this work, we test that model using the Chandra archival sample of 49 early-type galaxies from \citet{Lakhchaura2018}. We consider relationships between stellar velocity dispersion and entropy profile slope, multiphase gas extent, and the ratio of cooling time to freefall time. We also define subsamples based on data quality and entropy profile properties that clarify those relationships and enable more specific tests of the model predictions. We find that the atmospheric properties of early-type galaxies generally align with the predictions of the \citet{Voit2019} model, in that galaxies with greater stellar velocity dispersion tend to have radial profiles of pressure, gas density, and entropy with steeper slopes and less extended multiphase gas. Quantitative agreement with the model predictions improves when the sample is restricted to have low central entropy and stellar velocity dispersion of between 220 and $300\mathrm{~km~s^{-1}}$. \\
\end{abstract}

\section{Introduction}
Early-type galaxies, encompassing both elliptical and lenticular galaxies, are characterized by their elliptical shapes, older stellar populations, and lack of significant active star formation. Star formation in galaxies occurs when there is sufficient molecular gas to form stars and proceeds until the molecular gas supply runs out, either through stars forming more rapidly than the molecular gas can accumulate or the galaxy preventing further accumulation of molecular gas. It follows then, because little star formation is observed in early-type galaxies, that those galaxies must be preventing molecular gas from accumulating. Molecular gas can accumulate in galaxies via cold streams (e.g. \citealt[][]{Keres2005,Keres2009,Dekel2009}), cooling flows (e.g. \citealt[][]{White1991,Fabian1994,McnamaraNulsen2007,McnamaraNulsen2012,Werner2019}), stellar mass loss (e.g. \citealt[][]{Mathews2003,Leitner2011,VoitDonahue2011}), or perhaps tidal interactions with other galaxies. Therefore, feedback processes in early-type galaxies must limit those gas sources.

The effects of feedback processes on galactic atmospheres can be probed through observations of the hot X-ray emitting gas. One way feedback alters the rate at which gas flows into a galaxy is by affecting the cooling time of its atmospheric gas. For example, adding heat to a galactic atmosphere raises its entropy. If the heating due to feedback is gradual compared to the time it takes for the heated gas to expand within the gravitational potential, the temperature of the atmosphere may not change much while its density declines and its cooling time lengthens. Entropy is therefore the preferred quantity for investigating feedback processes and is represented in this paper by the entropy index $K\equiv kT n_e^{-2/3}$, where $kT$ is gas temperature and $n_e$ is electron density. We define the cooling time ($t_{\mathrm{cool}}$) of the gas to be the time needed for gas at temperature $T$ to radiate an energy $3kT/2$ per particle. For an atmosphere near hydrostatic equilibrium, the sound-crossing time required for heating to drive expansion is similar to the freefall time $t_{\mathrm{ff}} = (2r/g)^{1/2}$, defined with respect to the local gravitational acceleration $g$ at galactocentric radius $r$.  A galactic potential well serves as an entropy sorting device, causing lower entropy gas to sink, while higher entropy gas rises to larger radii. The atmosphere's lowest entropy gas therefore collects near the center, where it is the densest and brightest X-ray emitting gas in the galaxy.

\citet{Voit2015b} examined the properties of a sample of 10 massive elliptical galaxies previously studied by \citet[][]{Werner2012,Werner2014} and showed that the entropy profile slopes of early-type galaxies correlate with the presence of multiphase gas in that small sample. The gas entropy levels of those galaxies are similar inside the central $\sim$ 2 kpc, but their entropy profile slopes outside of $\sim$ 2 kpc differ. Galaxies with extended multiphase gas exhibit entropy profiles with $K\propto r^{2/3}$ at $\sim$1--10 kpc while galaxies with no extended multiphase gas (hereafter referred to as single phase galaxies) exhibit steeper entropy profiles, with $K\propto r$ at $\sim$1--10 kpc. 

\citet{Voit2015b} hypothesized that the differences in the entropy profile slope could be due to supernova (SN) type Ia heating that sweeps gas ejected by the old stellar population out of a single phase galaxy into an extended gaseous halo. They also found that the velocity dispersions of the galaxies with extended multiphase gas were $\sigma_v\leq255~\mathrm{km~s^{-1}}$ while galaxies with no extended multiphase gas had $\sigma_v\geq263~\mathrm{km~s^{-1}}$.  How a black hole interacts with a galactic atmosphere, as reflected by its entropy profile slope and multiphase gas extent, therefore appears to be related to the galaxy's velocity dispersion.

\citet{Lakhchaura2018} explored the relationship between entropy profile slope and multiphase gas extent in a larger archival sample ($\sim$50 galaxies) but did not report evidence for a relationship. However, \citet{Lakhchaura2018} did find evidence for a relationship between the average behavior of entropy profiles and the atmospheric ratio of cooling time to freefall time. \citet{Babyk2018} explored the relationship between entropy profile slope and velocity dispersion for an archival sample of 40 early-type galaxies (and 110 clusters). They did not report evidence for a relationship between entropy profile slope and velocity dispersion but did find some evidence for a relationship between entropy profile slope and atmospheric temperature.  

\citet{Voit2019} modeled the coupling between supernova sweeping of stellar ejecta, the confining circumgalactic medium (CGM) pressure, and bipolar kinetic feedback fueled by accretion of cooling gas onto the central black hole, and showed how it forms a ``black hole feedback valve." They presented an analytic model that predicts a simple relationship between a galaxy's stellar velocity dispersion and its entropy profile slope, based on how SN~Ia heating affects a galactic atmosphere and the locations where multiphase gas tends to form. The model is informed by both numerical simulations and observations and predicts the entropy profile slope within the radial range ($\sim$ 1--10 kpc) in which SN~Ia heating is capable of exceeding radiative cooling.  If SN~Ia heating does exceed radiative cooling there, then the entropy profile slope of the atmosphere in that region is determined by the ratio of the specific thermal energy of the ejected stellar gas to the depth of the galactic potential well, as long as the outflow that heating drives is subsonic.

The structure of this paper is as follows. Section \ref{methods} describes our sample selection and data analysis procedures. Section \ref{discussion} compares the model predictions of \citet{Voit2019} with those observations.  Section \ref{conclusions} concludes by discussing how this work relates to the current understanding of precipitation-driven feedback in massive galaxies. 

\section{Methods}\label{methods}
\subsection{Sample Description}

Our primary goal in this work is to determine whether the model predictions from \citet{Voit2019} are consistent with the observed relationships between stellar velocity dispersion and atmospheric entropy profile slope. Making such a comparison requires observations with sufficient resolution to measure an entropy profile slope, so we need to use a sample of early-type galaxies with accurate entropy profiles and velocity dispersion measurements to test the model's predictions.

\begin{figure*}
    \centering
    \includegraphics[width=\textwidth]{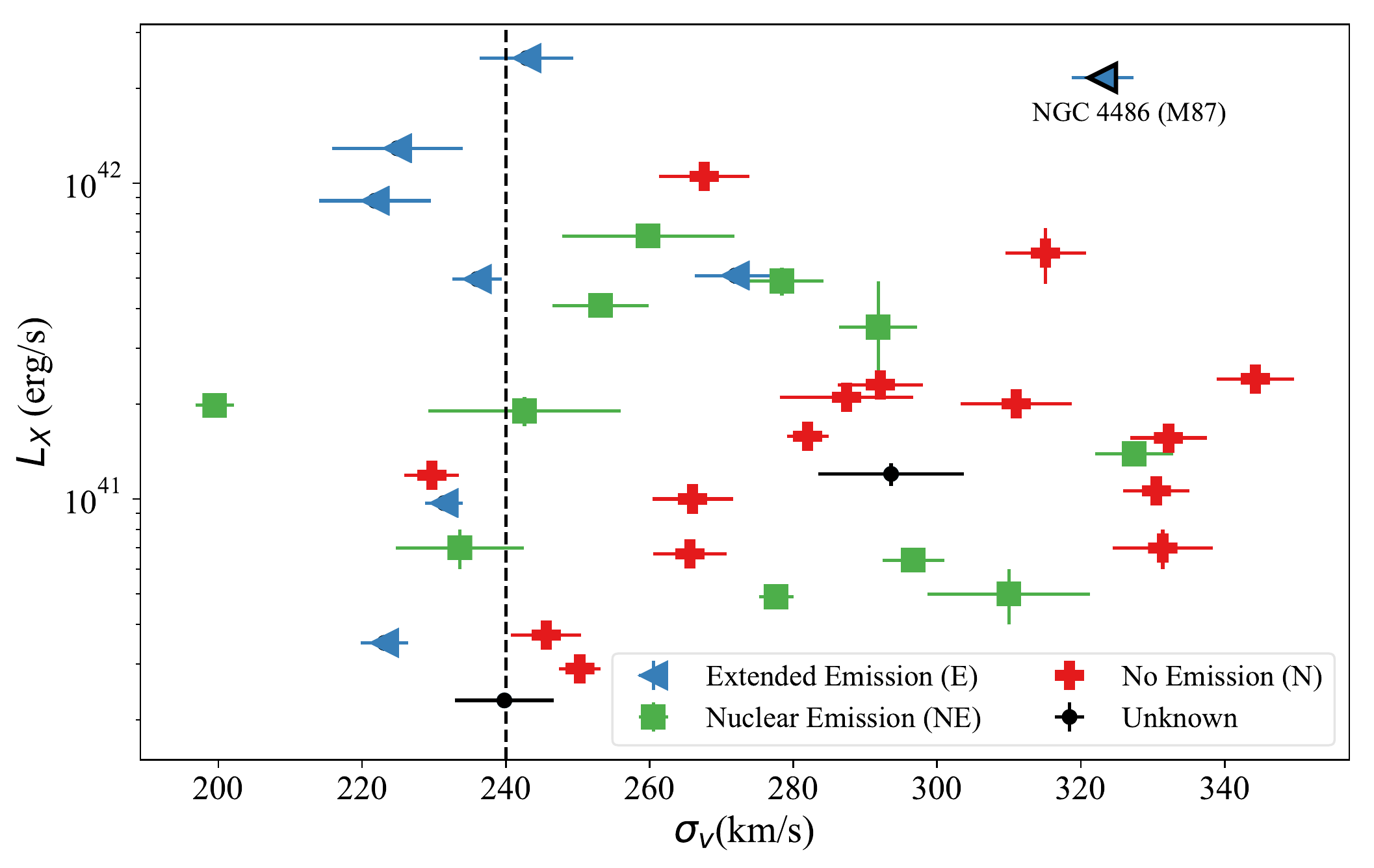}
    \caption{\textbf{Stellar velocity dispersion ($\sigma_v$) vs. X-ray luminosity ($L_\mathrm{X}$) from within the central 10~kpc of an early-type galaxy.} The figure shows our high-quality (HQ) subsample of galaxies from \citet{Lakhchaura2018} that have data quality sufficient for us to measure an accurate entropy profile slope at 1--10~kpc. Blue triangles represent galaxies with extended multiphase gas (E), red crosses represent galaxies with no extended multiphase gas (N), green squares represent galaxies with multiphase gas contained within 2 kpc (NE), and black dots represent galaxies without a multiphase gas classification from \citet{Lakhchaura2018}. A vertical dashed line indicates the velocity dispersion (240 $\mathrm{km~s^{-1}}$) that the \citet{Voit2019} model predicts should correspond to the critical entropy profile slope of $K \propto r^{2/3}$.
    \label{fig:sigmav_LX} }
\end{figure*}

The galaxies explored in this work were drawn from an analysis by \citet{Lakhchaura2018} of 49 nearby, X-ray and optically bright, elliptical galaxies with archival Chandra data. We adopted the radial profile measurements of electron density $n_{\mathrm{e}}$, temperature $kT$, specific entropy $K$, and the cooling time to freefall time ratio $t_{\mathrm{c}}/t_{\mathrm{ff}}$ from \citet{Lakhchaura2018}, as well as their classification scheme for the extent of multiphase gas. However, we restricted our analysis to a high-quality (HQ) subsample with a spatial resolution sufficient to fit a three-parameter entropy profile model in the 1--10 kpc range (see \S \ref{sample_selection} for details). 

Table \ref{tab:galaxy_params} lists the properties of the galaxies in the HQ subsample. According to the \citet{Lakhchaura2018} classification scheme, extended-emission (E) galaxies have optical line emission extending to more than 2~kpc from the center, nuclear-emission (NE) galaxies have line emission confined to the central 2~kpc, no-emission (N) galaxies have no detectable optical line emission, and U galaxies have unknown line-emission characteristics. The $\sigma_v$ values in Table \ref{tab:galaxy_params} are from the HYPERLEDA database.\footnote{\url{http://leda.univ-lyon1.fr}}  The cooling time $t_{\rm cool}$ is calculated from the X-ray observations assuming solar metallicity, and the estimated freefall time is taken to be $t_{\rm ff} = \sigma_v / r$.

Many of the emission-line classifications in \citet{Lakhchaura2018} were based on long-slit spectroscopy collected at the SOAR and Apache Point (APO) observatories, to be described in papers still in preparation. The SOAR observations come from the Goodman High-Throughput Spectrograph using the $1\farcs68$ slit and the 600 l/mm grating.  They cover $\sim 2600$~\AA, with an approximate spectral resolution of $R \sim 800$ at a central wavelength of 6000~\AA. The APO observations come from the Dual Imaging Spectrograph using the 2" slit and the R1200/B1200 grating. They cover $\sim 1200$ \AA, with an approximate spectral resolution of $R \sim 900$ at a central wavelength of 6600 \AA.  

The \citet{Lakhchaura2018} classifications should not be considered definitive, because the data sets on which they are based are inhomogeneous, with differing sensitivities and equivalent width limits. Future observations of greater sensitivity may therefore shift galaxies from the N class into the NE class, or from the NE class into the E class. However, we are retaining those classifications here, with one exception,\footnote{We have corrected the \citet{Lakhchaura2018} classification of IC 4296, shifting it from NE to E, for reasons described in \citet{Frisbie2020}.} so that the results of this paper can be compared directly with those presented in \citet{Lakhchaura2018}.

Figure \ref{fig:sigmav_LX} shows the relationships between X-ray luminosity, stellar velocity dispersion, and multiphase gas characteristics for galaxies in the HQ subsample. \citet{Voit2019} argued that the atmospheric properties of such galaxies should change above $\sigma_v=240\,\mathrm{km~s^{-1}}$, at which the analytical model they present predicts an entropy profile $K \propto r^{2/3}$ for reasons outlined in \S \ref{analytic_model_deriv}. In the HQ subsample, extended multiphase gas appears to be more common among galaxies with $\sigma_v \lesssim 240\,\mathrm{km~s^{-1}}$ than among those with $\sigma_v > 250\,\mathrm{km~s^{-1}}$. Figure \ref{fig:sigmav_LX} also shows  that galaxies with multiphase gas confined to the inner $\sim2$ kpc are represented across the full range of $\sigma_v$. The most notable exception to the tendency for galaxies with greater $\sigma_v$ to lack extended multiphase gas is M87 (also known as NGC 4486), which has an H$\alpha$ nebula extending several kiloparsecs from the center \cite[e.g.,][]{Sparks2004ApJ...607..294S,Boselli2019A&A...623A..52B}. It is the only galaxy in the upper right of Figure \ref{fig:sigmav_LX}, it resides in one of the most massive halos in the sample, and it has one of the greatest X-ray luminosities within 10~kpc. The large luminosity from that inner region results from high external gas pressure, applied by overlying gas in this halo. The atmospheric characteristics of M87 are therefore representative of the entire massive halo rather than the potential well of this single galaxy. 

Apart from M87, the upper envelope of $L_\mathrm{X}$ in Figure \ref{fig:sigmav_LX} exhibits a decline with increasing velocity dispersion. This feature is consistent with the hypothesis of \citealt[][]{Voit2015b,Voit2019}, which predicts that galaxies with greater velocity dispersions should have steeper entropy profile slopes. A steeper entropy profile slope means that the typical electron density within 10~kpc (the aperture for measuring $L_\mathrm{X}$) is generally smaller than it would be for a shallower entropy profile slope.

Section \ref{sample_selection} will discuss some smaller subsets of the \citet{Lakhchaura2018} sample that help to further refine our tests of the model predictions.

\begin{deluxetable*}{lcccccccc}
\tablecaption{
\label{tab:galaxy_params} \textbf{Properties of the HQ Subsample of Galaxies from \citet{Lakhchaura2018}}}

\setlength{\tabcolsep}{2pt}
\tablehead{
\colhead{Galaxy} &  \colhead{$z$}&  \colhead{$D$} &  \colhead{$\sigma_v$}  &  \colhead{$L_\mathrm{X}$} &  \colhead{Emission} &  \colhead{$\alpha_K$}  &  \colhead{min($t_{\rm c}/t_{\rm ff}$)}  &  \colhead{$K_0$}  \\ 
\colhead{} & \colhead{} & \colhead{(Mpc)} & \colhead{($\mathrm{km \, s^{-1}}$)}  & \colhead{($10^{42}\mathrm{erg \, s^{-1})}$} &  \colhead{Extent} & \colhead{} & \colhead{} &  \colhead{($\mathrm{keV \, cm^{2}}$)}}

\colnumbers \startdata
IC1860 &  0.0229 &  95.75 & $259.8 \pm 12.0$ &  $0.68 \pm 0.050$ & NE\tablenotemark{a} & $0.66 \pm 0.24$ &  $18.42 \pm 2.40$ &  $5.92 \pm 0.97$ \\
IC4296 &  0.0124 &  47.31 &    $327.4 \pm         5.4$ &  $0.14 \pm         0.003$ &         NE\tablenotemark{a} &     $1.23 \pm             0.11$ &          $11.63 \pm                 0.78$ &  $0.69 \pm         0.20$ \\
IC4765 &  0.0150 &  59.52 &    $278.4 \pm         5.8$ &  $0.49 \pm         0.050$ &         NE\tablenotemark{a} &     $0.88 \pm             0.22$ &          $11.01 \pm                 1.20$ &  $1.82 \pm         0.72$ \\
NGC315 &  0.0164 &  56.01 &    $293.6 \pm        10.1$ &  $0.12 \pm         0.010$ &          U &     $0.83 \pm             0.24$ &          $20.02 \pm                 4.65$ &  $0.91 \pm         0.96$ \\
NGC410 &  0.0176 &  66.00 &    $291.8 \pm         5.4$ &  $0.35 \pm         0.140$ &         NE\tablenotemark{b} &     $0.76 \pm             0.35$ &          $27.30 \pm                 5.22$ &  $4.93 \pm         1.36$ \\
NGC499 &  0.0147 &  60.74 &    $253.2 \pm         6.7$ &  $0.41 \pm         0.030$ &         NE\tablenotemark{b} &     $0.74 \pm             0.24$ &          $34.18 \pm                 3.19$ &  $6.90 \pm         4.16$ \\
NGC507 &  0.0164 &  59.83 &    $292.1 \pm         5.9$ &  $0.23 \pm         0.020$ &          N\tablenotemark{b} &     $0.80 \pm             0.34$ &          $30.15 \pm                 5.61$ &  $7.14 \pm         3.38$ \\
NGC533 &  0.0184 &  61.58 &    $271.9 \pm         5.6$ &  $0.51 \pm         0.030$ &          E\tablenotemark{a,d} &     $0.88 \pm             0.14$ &          $12.28 \pm                 3.75$ &  $1.74 \pm         0.36$ \\
NGC708 &  0.0162 &  64.19 &    $221.8 \pm         7.8$ &  $0.88 \pm         0.020$ &          E\tablenotemark{c} &     $0.64 \pm             0.06$ &          $12.04 \pm                 0.29$ &  $5.38 \pm         0.15$ \\
NGC741 &  0.0186 &  64.39 &    $287.4 \pm         9.3$ &  $0.21 \pm         0.010$ &          N\tablenotemark{a,b} &     $0.93 \pm             0.09$ &          $19.16 \pm                 0.73$ &  $2.57 \pm         0.39$ \\
NGC777 &  0.0167 &  58.08 &    $315.1 \pm         5.6$ &  $0.60 \pm         0.120$ &          N\tablenotemark{b} &     $0.59 \pm             0.24$ &          $24.11 \pm                 3.01$ &  $5.22 \pm         1.01$ \\
NGC1316 &  0.0059 &  19.25 &    $223.1 \pm         3.3$ &  $0.04 \pm         0.002$ &          E\tablenotemark{a} &     $0.72 \pm             0.25$ &          $32.57 \pm                 6.72$ &  $0.58 \pm         0.61$ \\
NGC1399 &  0.0048 &  17.75 &    $332.2 \pm         5.3$ &  $0.16 \pm         0.004$ &          N\tablenotemark{a} &     $0.94 \pm             0.03$ &          $26.05 \pm                 0.40$ &  $0.89 \pm         0.11$ \\
NGC1404 &  0.0065 &  19.18 &    $229.7 \pm         3.8$ &  $0.12 \pm         0.001$ &          N\tablenotemark{a} &     $0.80 \pm             0.03$ &          $20.23 \pm                 0.51$ &  $0.70 \pm         0.04$ \\
NGC1407 &  0.0060 &  23.27 &    $265.6 \pm         5.1$ &  $0.07 \pm         0.003$ &          N\tablenotemark{a} &     $0.83 \pm             0.06$ &          $41.92 \pm                 1.67$ &  $4.13 \pm         0.25$ \\
NGC1521 &  0.0140 &  50.93 &    $233.6 \pm         8.9$ &  $0.07 \pm         0.010$ &         NE\tablenotemark{a} &     $0.38 \pm             0.39$ &          $20.90 \pm                 5.21$ &  $0.96 \pm         0.83$ \\
NGC1600 &  0.0158 &  45.77 &    $331.4 \pm         7.0$ &  $0.07 \pm         0.010$ &          N\tablenotemark{a} &     $0.72 \pm             0.18$ &          $42.60 \pm                 7.16$ &  $5.07 \pm         0.39$ \\
NGC2300 &  0.0064 &  41.45 &    $266.0 \pm         5.6$ &  $0.10 \pm         0.010$ &          N\tablenotemark{b} &     $0.91 \pm             0.13$ &          $26.09 \pm                 1.27$ &  $4.18 \pm         0.43$ \\
NGC2305 &  0.0113 &  47.88 &    $242.6 \pm        13.4$ &  $0.19 \pm         0.020$ &         NE\tablenotemark{a} &     $0.70 \pm             0.25$ &          $20.22 \pm                 3.36$ &  $1.54 \pm         0.58$ \\
NGC3091 &  0.0122 &  48.32 &    $311.0 \pm         7.7$ &  $0.20 \pm         0.020$ &          N\tablenotemark{a} &     $0.40 \pm             0.10$ &          $30.74 \pm                 4.16$ &  $3.48 \pm         2.54$ \\
NGC3923 &  0.0058 &  20.97 &    $245.6 \pm         4.9$ &  $0.04 \pm         0.001$ &          N\tablenotemark{a} &     $0.92 \pm             0.12$ &          $21.98 \pm                 1.41$ &  $1.34 \pm         0.12$ \\
NGC4073 &  0.0197 &  60.08 &    $267.6 \pm         6.3$ &  $1.05 \pm         0.050$ &          N\tablenotemark{a} &     $0.61 \pm             0.20$ &          $32.22 \pm                 2.92$ &  $8.30 \pm         1.18$ \\
NGC4125 &  0.0045 &  21.41 &    $239.8 \pm         6.9$ &  $0.02 \pm         0.001$ &          U &     $0.13 \pm             0.45$ &          $28.22 \pm                12.35$ &  $1.43 \pm         1.40$ \\
NGC4261 &  0.0073 &  29.58 &    $296.7 \pm         4.3$ &  $0.06 \pm         0.003$ &         NE\tablenotemark{a,b} &     $1.16 \pm             0.06$ &          $14.17 \pm                 1.47$ &  $0.52 \pm         0.08$ \\
NGC4374 &  0.0033 &  16.68 &    $277.6 \pm         2.4$ &  $0.05 \pm         0.002$ &         NE\tablenotemark{a} &     $1.18 \pm             0.14$ &          $25.04 \pm                 6.58$ &  $1.86 \pm         0.19$ \\
NGC4406 &  0.0006 &  16.08 &    $231.4 \pm         2.6$ &  $0.10 \pm         0.004$ &          E\tablenotemark{c} &     $0.54 \pm             0.14$ &          $26.28 \pm                 1.60$ &  $5.21 \pm         3.26$ \\
NGC4472 &  0.0032 &  15.82 &    $282.0 \pm         2.9$ &  $0.16 \pm         0.001$ &          N\tablenotemark{a} &     $0.96 \pm             0.02$ &          $26.80 \pm                 0.23$ &  $1.17 \pm         0.05$ \\
NGC4486 &  0.0042 &  16.56 &    $323.0 \pm         4.3$ &  $2.16 \pm         0.004$ &          E\tablenotemark{c} &     $0.61 \pm             0.01$ &          $22.73 \pm                 0.27$ &  $3.00 \pm         0.10$ \\
NGC4552 &  0.0009 &  15.97 &    $250.3 \pm         2.9$ &  $0.03 \pm         0.001$ &          N\tablenotemark{a} &     $0.95 \pm             0.09$ &          $11.35 \pm                 0.63$ &  $2.23 \pm         0.11$ \\
NGC4636 &  0.0031 &  15.96 &    $199.5 \pm         2.7$ &  $0.20 \pm         0.002$ &         NE\tablenotemark{a} &     $1.00 \pm             0.03$ &          $10.79 \pm                 0.36$ &  $1.89 \pm         0.08$ \\
NGC4649 &  0.0034 &  16.55 &    $330.5 \pm         4.6$ &  $0.11 \pm         0.002$ &          N\tablenotemark{a} &     $1.00 \pm             0.02$ &          $22.63 \pm                 0.35$ &  $1.49 \pm         0.02$ \\
NGC4696 &  0.0098 &  37.48 &    $242.9 \pm         6.5$ &  $2.49 \pm         0.010$ &          E\tablenotemark{c} &     $0.69 \pm             0.01$ &           $4.73 \pm                 0.03$ &  $2.24 \pm         0.07$ \\
NGC4782 &  0.0133 &  48.63 &    $310.0 \pm        11.3$ &  $0.05 \pm         0.010$ &         NE\tablenotemark{a,c} &     $0.59 \pm             0.26$ &          $18.94 \pm                 9.92$ &  $4.30 \pm         2.62$ \\
NGC5044 &  0.0090 &  35.75 &    $224.9 \pm         9.1$ &  $1.29 \pm         0.010$ &          E\tablenotemark{a} &     $0.56 \pm             0.03$ &          $ 5.75 \pm                 0.14$ &  $0.08 \pm         0.12$ \\
NGC5419 &  0.0139 &  50.87 &    $344.3 \pm         5.4$ &  $0.24 \pm         0.020$ &          N\tablenotemark{a} &     $1.19 \pm             0.28$ &          $17.30 \pm                 2.21$ &  $1.38 \pm         0.70$ \\
NGC5813 &  0.0064 &  29.23 &    $236.0 \pm         3.4$ &  $0.50 \pm         0.003$ &          E\tablenotemark{a} &     $0.51 \pm             0.02$ &          $12.20 \pm                 0.26$ &  $3.44 \pm         0.13$ 
\enddata
    \tablenotetext{a}{Classification based on SOAR long-slit spectroscopy (T. Connor et al., in preparation).}
    \tablenotetext{b}{Classification based on APO long-slit spectroscopy (M. Sun et al., in preparation).}
    \tablenotetext{c}{See \citet{Lakhchaura2018} for classification data source.}
    \tablenotetext{d}{H$\alpha$ emission-line map from \citet{Hamer_2016MNRAS.460.1758H}}
\end{deluxetable*}

\subsection{Model Predictions}\label{analytic_model_deriv}

\citet{Voit2019} presented a simple analytical prediction for the relationship between entropy profile slope ($\alpha_K \equiv d \ln K / d \ln r$) and stellar velocity dispersion ($\sigma_v$) in an early-type galaxy. The basic model assumes that the galaxy's stellar mass distribution can be approximated by a singular isothermal sphere (constant circular velocity $v_c$) with a one-dimensional velocity dispersion $\sigma_v=v_c/\sqrt{2}$. If SN~Ia heating locally dominates radiative cooling and black hole feedback, then it will drive an outflow that is subsonic within the galaxy, and therefore close to hydrostatic equilibrium. The radial power-law slopes of pressure, density, and entropy in such an outflow depend on how the specific energy of the gas compares with the gravitational potential. If gravity is insignificant, those power-law slopes will be shallow, but if the gravitational potential energy is comparable to the thermal energy, then those slopes will be steep. Consequently, outflows driven by stellar heating from galaxies with deeper central potentials, corresponding to larger $\sigma_v$, should have profiles of pressure, density, and entropy that are more centrally concentrated and focus radiative cooling more tightly around the central black hole.

Combining the contributions to the entropy profile from supernova energy, stellar orbital energy, and gravitational potential energy gives the following quantitative prediction for the relationship between $\alpha_K$ and $v_c$:
\begin{equation}\label{eq:alpha_sigma}
    \alpha_K\approx\frac{5}{3}\Big(\frac{\epsilon_*}{v_c^{2}}-\frac{1}{4}\Big)^{-1} 
    \; \; ,
\end{equation}
where $\epsilon_*$ is the mean specific energy of the gas coming from stars, including SN~Ia heating, which amounts to $\approx 2$~keV per particle for reasons outlined in \citet{Voit2019}. Therefore, the structure of the galaxy's atmosphere at 1--10~kpc depends strongly on $\epsilon_*/v_c^{2}$, which is equivalent to $\epsilon_*/2 \sigma_v^2$. Section \ref{analytic_prediction_comp} compares this prediction with the HQ subsample.

\citet{Voit2019} also presented a slightly more realistic form of the basic model, assuming that the galaxy's halo has a Navarro-Frenk-White (NFW) density profile and the stellar mass density follows a modified Einasto profile. Numerical integration of the more realistic model shows that the basic model overpredicts the entropy profile slope for $\sigma_v~>~300$ $\mathrm{km~s^{-1}}$.  We compare that modification of the basic model with the HQ subsample in in \S \ref{num_int_comp}.  

\subsection{Entropy Profile Measurements} \label{K_profile_fitting}

Equation \ref{eq:alpha_sigma} is based on a steady, pressure bounded, subsonic outflow solution, heated only by SN~Ia in an isothermal potential, and predicts a constant value of $\alpha_K$  in the region where those conditions apply. Because we want to test that prediction for the relationship between $\alpha_K$ and $\sigma_v$, we limit the range over which we fit $\alpha_K$ to the radial range that is affected as little as possible by other heating processes. 

If feedback from the active galactic nucleus (AGN) is as powerful as in NGC 4261 and IC 4296 (see \citealt{Frisbie2020}), it typically deposits its energy rather far from the center ($r>10$~kpc) because its jets drill through the hot gas, allowing stellar heating to dominate at smaller radii.  However, a fraction of that AGN energy output might couple to gas closer to the AGN in some galaxies, resulting in a flattening or even an inversion of the entropy profile near 1 kpc. Therefore, we limit our entropy profile slope measurements to 1--10 kpc to get a ``clean'' measure of $\alpha_K$ where stellar processes are most likely to dominate.

While a few of the galaxies in our sample have entropy profiles that resemble a pure power law \citep{Frisbie2020}, most have an excess of entropy over a pure power law in the central kiloparsec. Therefore, we have adapted the functional form used by  \citet{Donahue2005,Donahue2006} and \citet{Cavagnolo2009} to a radial range more appropriate for individual galaxies instead of galaxy clusters:
\begin{equation}
    \label{eq:K_profile_fit}
    K(r)=K_0+K_{10} \left(\frac{r}{10~ \mathrm{kpc}}\right)^{\alpha_{K}},
\end{equation}
where $K_0$ is a constant core entropy, $K_{10}$ is the normalization of the power-law component at a radius of 10 kpc, and $\alpha_{K}$ is the best-fitting power-law slope. 

\begin{figure}[t!]
    \centering
    \includegraphics[width=0.45\textwidth,trim={0.2in 0in 0in 0in}]{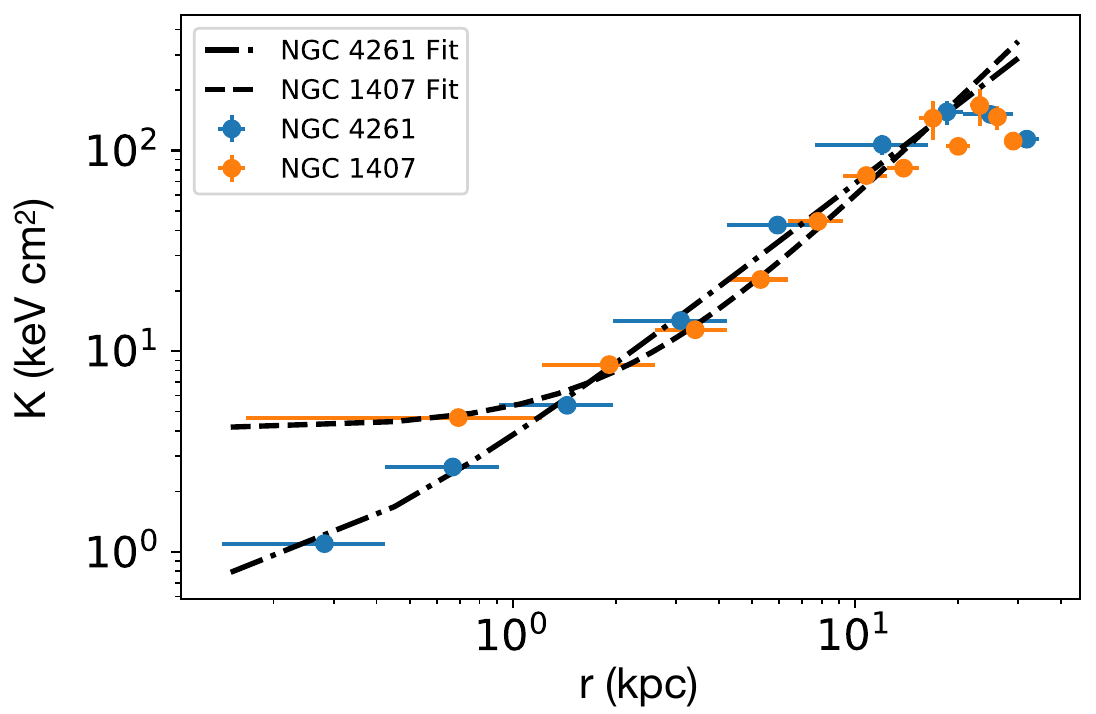}
    \caption{\textbf{Examples of entropy profile fits.} In NGC~4261 (blue points), the entropy profile is a nearly pure power law, so the best-fitting profile (dotted-dashed line) has a small value of $K_0$. The entropy profile of NGC~1407 (orange points) has a significantly greater value of $K_0$, reflected by the central flattening of its best-fitting entropy profile (dashed line).  
    \label{fig:K-r_fits}}
\end{figure}

We calculate the best-fit parameters using the Python package \texttt{emcee} \citep{Foreman-Mackey_emcee_2013PASP..125..306F}. We establish an initial broad expected range for the parameters in log space with $0<K_0<10^2$, $0<K_{10}<10^2$, and $0<\alpha_{K}<2$. Errors were determined from Markov Chain Monte Carlo (MCMC) contours in two dimensions ($16-84\%$). Figure \ref{fig:K-r_fits} shows two examples of entropy profile fits to high-quality data.
 
\begin{figure}[t!]
    \centering
    \includegraphics[trim={0in 0in 0in 0in},width=0.45\textwidth]{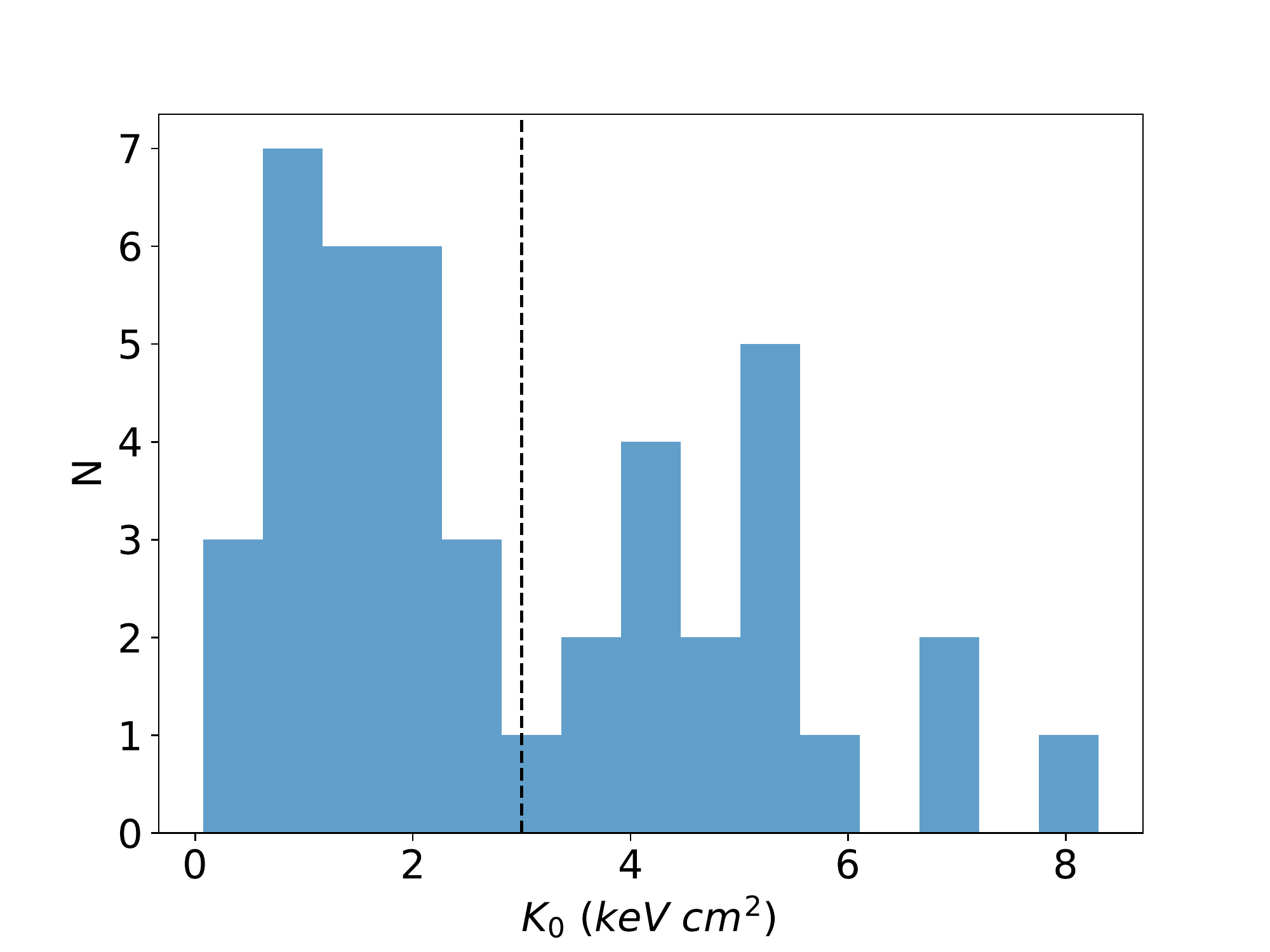}
    \caption{\textbf{Distribution of $\pmb{K_0}$ values for the HQ subsample.}  The histogram shows $K_0$ values obtained by fitting Equation \ref{eq:K_profile_fit} in the 1--10 kpc radial interval to galaxies with sufficient spatial resolution. The black dashed vertical line at $K_0=3~\mathrm{keV ~cm}^2$ represents the criterion used to define the low $K_0$ subset. 
    \label{fig:K0_hist}}
\end{figure}

Our entropy profile fits suggest that approximately half of the HQ subsample is clustered near $K_0 \sim 1-2 \, \mathrm{keV~cm}^2$, while the rest have $K_0 \gtrsim 3 \, \mathrm{keV~cm}^2$ (see Figure \ref{fig:K0_hist}). In the group with low $K_0$, the entropy profile at 1--10 kpc is close to a pure power law. This set of galaxies is therefore more suitable for testing the prediction represented in Equation \ref{eq:alpha_sigma}. 

The subset with greater $K_0$, by contrast, clearly deviates from the pure power-law entropy profile predicted by the basic analytical model. In those galaxies, central AGN heating may be producing an entropy floor at small radii, causing a flattening of the power-law profile that extends into the 1--10~kpc region. The best-fitting values of $\alpha_K$ in the high $K_0$ galaxies may still be representative of SN~Ia heating, but the measurements of $\alpha_K$ are not as clean because of a greater degeneracy between $K_0$ and $\alpha_K$ in the fitting procedure.

\subsection{Subset Selection}\label{sample_selection}

Our selection of the HQ subsample from the \citet{Lakhchaura2018} sample was intended to test the prediction described by Equation \ref{eq:alpha_sigma} as accurately as possible. Because the analysis requires a statistically significant measurement of the slope parameter $\alpha_K$, the HQ subsample includes only those galaxies from \citet{Lakhchaura2018} with sufficient spatial resolution, defined to be at least 4 radial bins of any width in the 1--10 kpc radial interval. There are 36 galaxies in the \citet{Lakhchaura2018} sample that fit this criterion, and those are the ones making up the HQ subsample listed in Table \ref{tab:galaxy_params}.  

Two further subsets of the \citet{Lakhchaura2018} sample are useful for the analysis in \S \ref{discussion}.  The first is selected based on $K_0$.  Applying the cut at $K_0 = 3 \, {\rm keV \, cm^2}$ shown in Figure \ref{fig:K0_hist} defines a low $K_0$ subset of 22 galaxies that provide a cleaner test of the model predictions.  A more restricted subset (hereafter referred to as the limited $\sigma_v$ subset) is based on $\sigma_v$.  It further limits the low $K_0$ subset to galaxies with $\sigma_v = 220$--$300 \,  \mathrm{km~s^{-1}}$. The rationale for focusing on that particular subset of 16 galaxies is discussed in Section \ref{analytic_prediction_comp}. 

\section{Discussion}\label{discussion}

\subsection{Tests of the Analytical Prediction}\label{analytic_prediction_comp}

The \citet{Voit2019} analytic model predicts a relationship for stellar velocity dispersion and entropy profile slope (Equation \ref{eq:alpha_sigma}) in the radial range where SN~Ia heating may be significant (within $\sim$1--10 kpc). Elevated central entropy ($K_0$) beyond 1 kpc suggests that the central AGN is more strongly coupled to the surrounding medium at small radii \citep{Prasad2020}. Therefore, SN~Ia heating may not be the only heating process at 1--10~kpc, in which case the simple analytical model does not strictly apply. Because the model presumes that AGN heating does not elevate the central entropy, we are motivated to investigate the relationship between velocity dispersion and entropy profile slope in the low $K_0$ subset as well as the HQ subsample.

\begin{table*}
\caption{Fits to the Observed Slope of the $\sigma_v$--$\alpha_K$ Relationship} 
\setlength{\tabcolsep}{4pt}
\begin{center}
\begin{tabular}{ccccccc}
\hline
\textbf{Sample} & $\alpha_{240}$ &  $\alpha_K^\prime$  &  \textbf{Reduced $\chi^2$} &  \textbf{Intrinsic Scatter} & \textbf{Number of Galaxies} \\  
\hline \hline
High Quality (HQ) & $0.70 \pm 0.15$ & $0.52\pm0.24$ & 1.03  & $0.18\pm0.02$ & 36 \\
low $K_0$ & $0.74 \pm 0.27$ &  $0.80\pm0.33$ & 1.05  & $0.22\pm0.03$ & 22 \\
limited $\sigma_v$  & $0.66 \pm 0.19$ & $1.80\pm0.51$ & 1.08  & $0.16\pm0.02$ & 16 \\
\hline \hline
\end{tabular}
\label{tab:fit_results}
\end{center}
\end{table*}

Table \ref{tab:fit_results} summarizes the results of our exploration of the relationship between velocity dispersion and entropy profile slope for all three subsets defined in \S \ref{sample_selection}. To quantify the potential relationship between $\alpha_{K}$ and $\sigma_v$, we expand that relationship to linear order around the critical velocity dispersion to obtain the fitting formula
\begin{equation}
    \alpha_K = \alpha_{240} + \alpha_K^\prime (\sigma_{240} - 1) 
\end{equation}
where $\alpha_{240}$ is the best fit to $\alpha_K$ at $\sigma_v = 240 \, {\rm km \, s^{-1}}$, $\alpha_K^\prime$ is the best-fitting slope of the $\sigma_v$--$\alpha_K$ relation in the vicinity of that point, and $\sigma_{240} \equiv \sigma_v / (240 \, {\rm km \, s^{-1}})$.  We determine the best fits using an ordinary least squares method \citep{Akritas1996}, allowing for intrinsic scatter in $\alpha_K$.   

\begin{figure*}
    \centering
    \includegraphics[width=\textwidth]{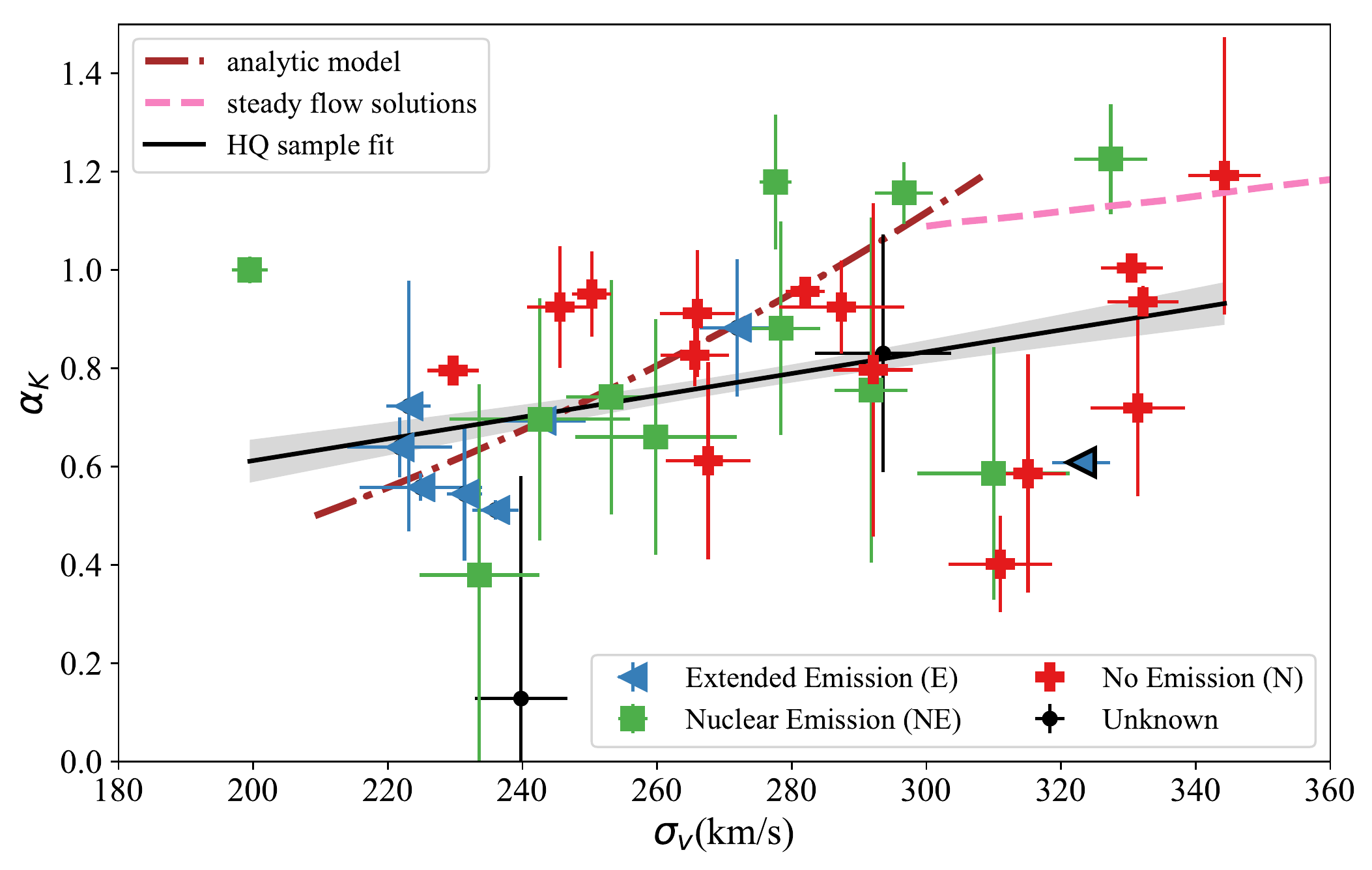}
    \caption{\textbf{Stellar velocity dispersion vs. entropy profile slope for the HQ subsample of galaxies.} 
    Point colors and shapes represent multiphase gas extent as classified by \citet{Lakhchaura2018}. Blue triangles are galaxies with extended multiphase gas (E), red crosses are galaxies with no extended multiphase gas (N), green squares are galaxies with multiphase gas contained within 2 kpc (NE), and black dots are galaxies without a gas extent classification. A black line shows the ordinary least squares fit to the data, with a gray band showing the $1\sigma$ error. A maroon dotted-dashed line represents the analytical prediction described by Equation \ref{eq:alpha_sigma}. A pink dashed line represents the steady flow solutions for $\sigma_v\gtrsim300~\mathrm{km~s^{-1}}$ from \citet{Voit2019}.
    \label{fig:sigma_v_alpha_all_galaxies}}
\end{figure*}

Figure \ref{fig:sigma_v_alpha_all_galaxies} illustrates our results for the HQ subsample. Our linear fit gives $\alpha_{240} = 0.70 \pm 0.15$, consistent with the analytical prediction, with a slope of $\alpha_K^\prime = 0.53\pm0.27$.  While that slope differs from zero by only about $2\sigma$, it represents tentative evidence for the predicted rise in $\alpha_K$ with $\sigma_v$, before the HQ subsample is further reduced.  Therefore, limiting the radial range of the entropy profile fit and requiring sufficient data resolution over that radial range, as we have done here, help to reduce some of the ambiguity in the $\sigma_v$--$\alpha_K$ relation present in previous work (e.g. \citealt{Babyk2018,Lakhchaura2018}).

\begin{figure*}
    \centering
    \includegraphics[width=\textwidth]{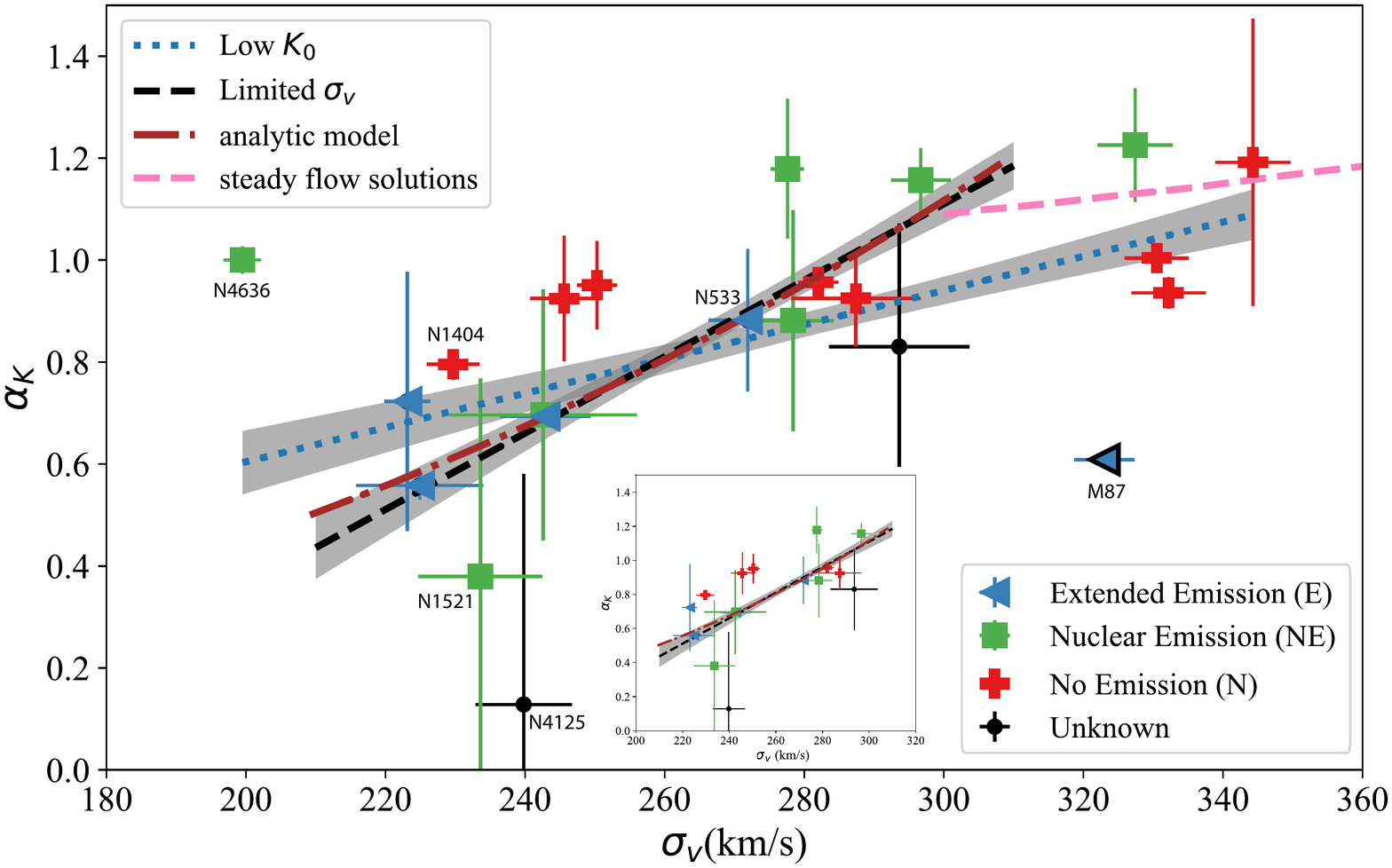}
    \caption{\textbf{Stellar velocity dispersion vs. entropy profile slope for the low $K_0$ and restricted $\sigma_v$ subsets.}  Points show the low $K_0$ subsample and are coded as in Figure \ref{fig:sigma_v_alpha_all_galaxies}. The dotted-dashed maroon and dashed pink lines are identical to the ones in Figure \ref{fig:sigma_v_alpha_all_galaxies}.  A blue dotted line shows the best fit to the low $K_0$ sample, and a black dashed line shows the best fit to the limited $\sigma_v$ sample, with gray bands representing $1\sigma$ errors. The inset shows just the restricted $\sigma_v$ subset. The labeled galaxies are those galaxies discussed in \S \ref{comments_on_indiv_galaxies}.}
    \label{fig:sigma_v_alpha_restricted}
\end{figure*}

However, the HQ sample contains some galaxies that the analytical model of \citet{Voit2019} was not designed to describe. The low $K_0$ subset excludes many of those galaxies (see \S \ref{sample_selection} for the rationale), and Figure \ref{fig:sigma_v_alpha_restricted} shows our fitting results for that subset. At $240 \, {\rm km \, s^{-1}}$, the best fit for $\alpha_K$ is essentially the same ($\alpha_{240} = 0.74 \pm 0.27$), but the slope of its dependence on $\sigma_v$ is steeper ($\alpha_K^\prime = 0.80\pm0.33$), providing stronger but not conclusive evidence for a positive $\sigma_v$--$\alpha_K$ correlation when the galaxies with elevated central entropy are removed.

Figure \ref{fig:sigma_v_alpha_restricted} also shows the best fit to the limited $\sigma_v$ subset defined in \S \ref{sample_selection}. This subset excludes the single galaxy with $\sigma_v <$ 220 $\mathrm{km~s^{-1}}$ (NGC 4636, see \S \ref{sec:N4636}) because the \citet{Voit2019} analytical model is not self-consistent for galaxies with $\sigma_v$ much less than $240 \, {\rm km \, s^{-1}}$.  The model assumes a steady outflow in which SN~Ia exceeds radiative cooling everywhere.  However, it predicts $\alpha_K < 2/3$ for $\sigma_v < 240 \, {\rm km \, s^{-1}}$, which leads to an inconsistency, because the ratio of SN~Ia heating to radiative cooling then declines with increasing radius for $\alpha_K < 2/3$ (see \citealt{Voit2019} for details). Consequently, such an atmosphere is prone to convective and thermal instabilities, because a cooling-dominated region sits on top of a heating-dominated region.  It is therefore unlikely to settle into a steady state like the one the model assumes.  Instead, it should experience episodic feedback, fueled by intermittent accretion of multiphase gas, causing its entropy profile to fluctuate within a range spanning $5 < t_{\rm cool} / t_{\rm ff} < 20$ (see \citealt{Prasad2020} for simulations demonstrating this behavior).

The limited $\sigma_v$ subset also excludes the 5 galaxies from the low $K_0$ subset that have $\sigma_v >$ 300 $\mathrm{km~s^{-1}}$.  There are two reasons for this restriction. First, the \citet{Voit2019} model assumes a singular isothermal sphere for the galaxy's stellar mass profile (see Section \ref{analytic_model_deriv}), which results in an overprediction of $\alpha_K$ at large $\sigma_v$, relative to numerical integrations of more realistic models (see the dashed pink line in Figures \ref{fig:sigma_v_alpha_all_galaxies}). Second, some galaxies in the sample, like M87, are central galaxies in galaxy groups or clusters, and thus are in potential wells considerably deeper than indicated by the central galaxy's stellar velocity dispersion.  Consequently, they have greater circumgalactic gas pressure that does not allow SN~Ia heating to drive the kind of outflow assumed by the analytic model. Applying an upper limit on stellar velocity dispersion therefore restricts the sample to galaxies most representative of the scenario the analytic model describes. 

For this limited $\sigma_v$ subset, we find best fits of $\alpha_{240} = 0.66 \pm 0.19$ and $\alpha_K^\prime = 1.80\pm0.51$, which is more than $\sim3\sigma$ away from a flat line. The criteria we have applied limit the subset to the galaxies most likely to follow the analytic model, so it is encouraging that the evidence for a relationship becomes stronger when the analysis focuses on galaxies with properties that are consistent with the analytical model's assumptions.  With these restrictions applied, the entropy profile slope appears to be much more closely connected to stellar velocity dispersion than indicated by previous works (e.g. \citealt{Babyk2018,Lakhchaura2018}).   Also, the analytical prediction of equation (\ref{eq:alpha_sigma}), which has no free parameters, appears to match the $\sigma_v$--$\alpha_K$ relation observed among low $K_0$ early-type galaxies with $220 \, {\rm km \, s^{-1}} \lesssim \sigma_v \lesssim 300 \, {\rm km \, s^{-1}}$

\subsection{Comparison to Numerical Predictions}\label{num_int_comp}

Taken at face value, equation (\ref{eq:alpha_sigma}) implies that $\alpha_K$ increases without bound as $\sigma_v$ rises and that it becomes infinite at a finite value of $\sigma_v$.  Before that can happen, one or more of the assumptions in the analytical model must break down.  The numerically integrated steady flow solutions presented in \citet{Voit2019} indicate that equation (\ref{eq:alpha_sigma}) indeed becomes increasingly invalid as $\sigma_v$ rises above $300 \, {\rm km \, s^{-1}}$.  That is because $\alpha_K$ at 1--10~kpc becomes increasingly sensitive to the effective radius of the galaxy's stellar mass distribution, which the simple analytical model does not account for. As a result, the numerically predicted entropy slope $\alpha_K$ levels off near 1.2, as indicated by the pink dashed lines in Figures \ref{fig:sigma_v_alpha_all_galaxies} and \ref{fig:sigma_v_alpha_restricted}.  In practice, precise predictions for $\alpha_K$ in this range of $\sigma_v$ require specification of the effective radius of the stellar mass distribution, introducing another galaxy parameter that is beyond the scope of this comparison.

Figure \ref{fig:sigma_v_alpha_restricted} shows that $\alpha_K$ in the low $K_0$ subset likewise levels off near 1.2 at $\sigma_v > 300 \, {\rm km \, s^{-1}}$.  Among the five galaxies with $\sigma_v > 300 \, {\rm km \, s^{-1}}$, four have $\alpha_K$ between 0.9 and 1.2.  The fifth is M87, which will be discussed separately in \S \ref{M87}.  The fact that the other four galaxies are close to the numerical model prediction but not the analytical prediction aligns with our rationale for truncating the low $K_0$ galaxies with $\sigma_v > 300 \, {\rm km \, s^{-1}}$ when defining the limited $\sigma_v$ subset.  However, additional observations of low $K_0$ galaxies with $\sigma_v > 300 \, {\rm km \, s^{-1}}$ are needed to demonstrate more conclusively that the $\sigma_v$--$\alpha_K$ relation flattens as predicted by numerical models in the limit of $\sigma_v > 300 \, {\rm km \, s^{-1}}$.

\subsection{Multiphase Gas Predictions}

\begin{figure*}
    \centering
    \includegraphics[width=\textwidth]{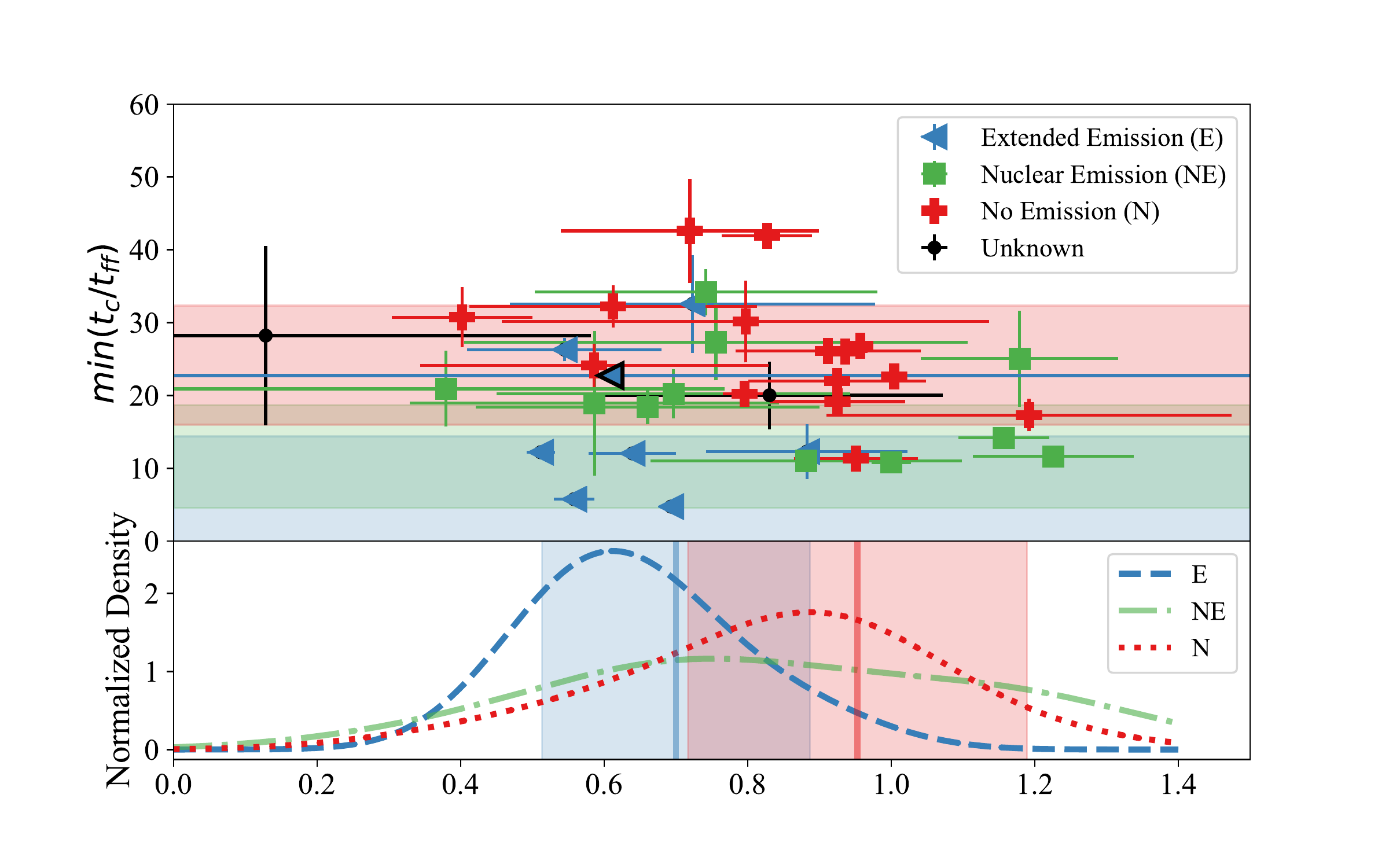}
    \caption{\textbf{Dependence of multiphase gas extent on $\alpha_K$ and min($t_{ \mathrm{cool}}/t_{ \mathrm{ff}}$) in the HQ subsample.} The top panel shows how the \citet{Lakhchaura2018} multiphase extent classification depends on both $\alpha_K$ and min($t_{ \mathrm{cool}}/t_{ \mathrm{ff}}$), with point colors and types as defined in Figure \ref{fig:sigma_v_alpha_all_galaxies}.  The bottom panel shows the distribution of $\alpha_K$ values for each multiphase class, smoothed using gaussian Kernel Density Estimation (KDE). A blue dashed line shows galaxies with extended multiphase gas (E).  A dotted-dashed green line shows galaxies with multiphase gas that does not extended beyond 2 kpc (NE).  A dotted red line shows galaxies with no extended multiphase gas (N). The solid vertical lines correspond to the emission classifications in color and are the predicted $\alpha_{\rm K}$ values calculated from Equation \ref{eq:alpha_sigma} using the average $\sigma_v$ values for E (blue line) and N (red line) galaxies, with shading corresponding to the predicted dispersion in slope.
    Bandwidths used for KDE were E: 0.115, NE: 0.192, and N: 0.152. 
    \label{fig:alphak_mintctff_all}}
\end{figure*}


\begin{figure*}
    \centering
    \includegraphics[width=\textwidth]{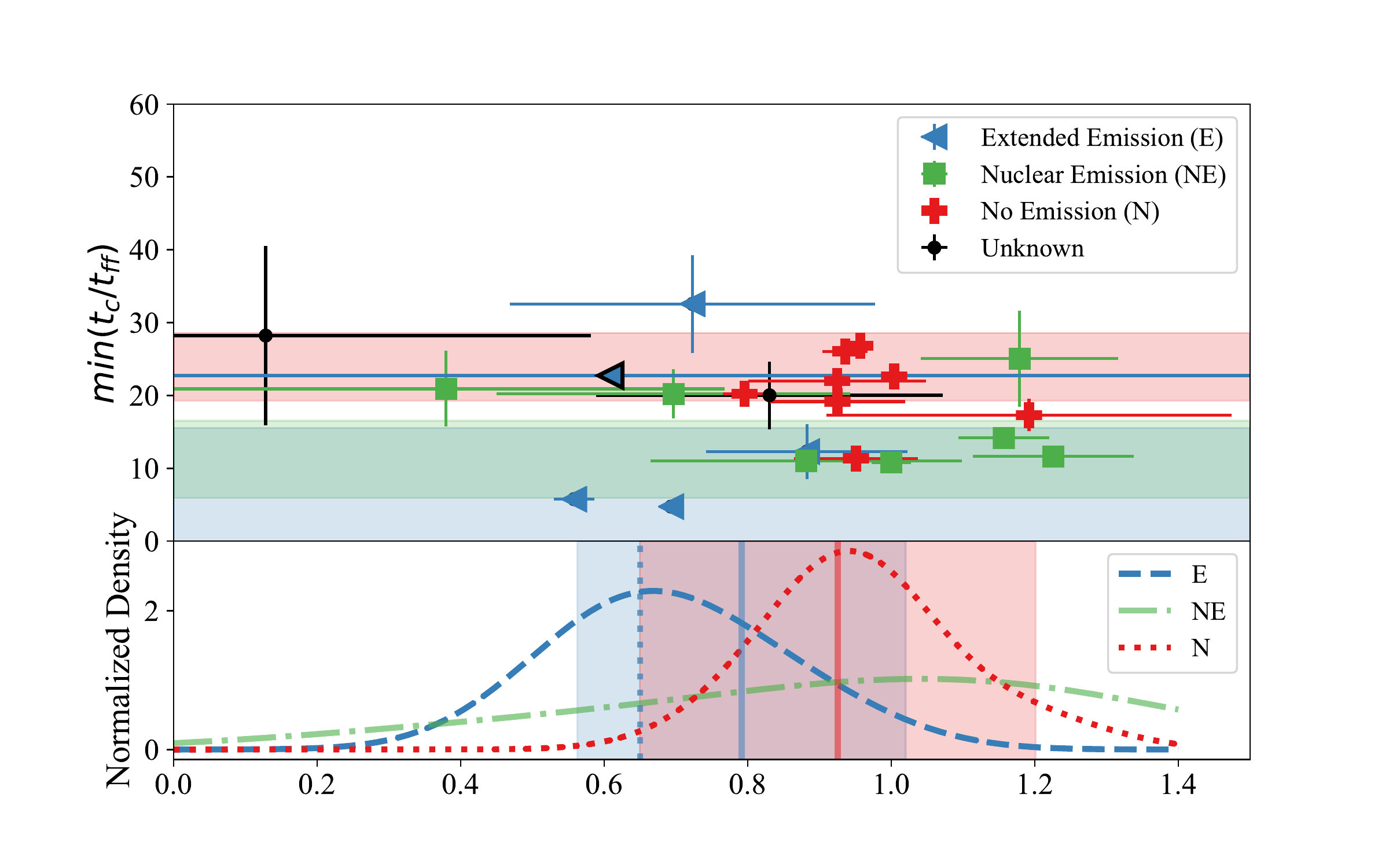}
    \caption{\textbf{Dependence of multiphase gas extent on $\alpha_K$ and min($t_{ \mathrm{cool}}/t_{\mathrm{ff}}$) in the low $\pmb{K_0}$ subset.} The top panel shows how the \citet{Lakhchaura2018} multiphase extent classification depends on both $\alpha_K$ and min($t_{ \mathrm{cool}}/t_{ \mathrm{ff}}$), with point colors and types as defined in Figure \ref{fig:sigma_v_alpha_all_galaxies}.  The bottom panel shows the distribution of $\alpha_K$ values for each multiphase class, smoothed using gaussian Kernel Density Estimation (KDE). The solid vertical lines are the predicted $\alpha_{\rm K}$ values calculated using the average $\sigma_v$ for E and N galaxies and Equation \ref{eq:alpha_sigma}, and the dashed vertical line is the predicted $\alpha_{\rm K}$ if M87 is removed from the multiphase galaxies. All figure elements have the same meanings as in Figure \ref{fig:alphak_mintctff_all}.  Bandwidths used for KDE were E: 0.132, NE: 0.278, and N: 0.110.
    \label{fig:alphak_mintctff_restricted}}
\end{figure*}

The black hole feedback valve model of \citet{Voit2019} also makes predictions for the extent of multiphase gas within an early-type galaxy.  It predicts that extended multiphase gas should be prevalent among early-type galaxies with $\sigma_v \lesssim 240 \, {\rm km \, s^{-1}}$, which should also have $\alpha_K \lesssim 2/3$ and $5 \lesssim \min(t_{\rm cool}/t_{\rm ff}) \lesssim 20$.  That prediction stems from the fact that $t_{\rm cool}/t_{\rm ff}$ declines with increasing radius for $\alpha_K \lesssim 2/3$, making the periphery of such a galaxy unstable to multiphase precipitation \cite[see, e.g.,][]{Voit2017,Voit2019}.  

It also predicts that multiphase gas should become less extended and increasingly concentrated toward a galaxy's center as $\sigma_v$ increases, as long as the circumgalactic gas pressure is low enough for SN~Ia heating to exceed radiative cooling within the galaxy.  There are two reason for this concentration of multiphase gas toward the center. First, the $t_{\rm cool}/t_{\rm ff}$ ratio of a galactic atmosphere with $\alpha_K > 2/3$ {\em increases} with increasing radius, meaning that the region of the atmosphere most prone to thermal instability is at small radii.  Second, the ratio of SN~Ia heating to radiative cooling within the galaxy also increases with radius for $\alpha_K > 2/3$, meaning that cooling is most likely to exceed heating at small radii.  However, these predictions for correlations of multiphase gas extent with $\sigma_v$ do not apply if the circumgalactic pressure is large enough for radiative cooling to exceed SN~Ia heating within the galaxy, as in the central galaxies of galaxy clusters.  In that case, AGN feedback is expected to maintain the central gas in a state with $5 \lesssim \min(t_{\rm cool}/t_{\rm ff}) \lesssim 20$ and $\alpha_K \approx 2/3$, which is marginally unstable to the production of extended multiphase gas.

Figures \ref{fig:alphak_mintctff_all} and \ref{fig:alphak_mintctff_restricted} present the relationships we find among multiphase gas extent, entropy profile slope at 1--10~kpc, and the minimum value of $t_{\rm cool}/t_{\rm ff}$ in the ambient galactic atmosphere.  In making those figures, we have employed Kernel Density Estimation (KDE) to produce continuous distribution functions that express the dependence of multiphase gas extent on $\alpha_K$. The KDE procedure convolves the discrete data with a smooth kernel function, in this case, a one-dimensional Gaussian with constant bandwidth. We chose KDE because it captures features of the distribution that could be masked by the choice of bin size in a histogram. 

We chose a Gaussian kernel because the data are relatively simple and one dimensional. Bandwidths were determined for each multiphase gas extent category by minimizing the mean integrated square error (MISE):
\begin{equation}
    {\rm MISE}(\hat{f}_{\rm kern}) = {\rm E} \left\{ \int [\hat{f}_{\rm kern}(x)-f(x)]^2 dx \right\}
    \; \; ,
\end{equation}
where $\hat{f}_{\rm kern}$ is the chosen kernel function, E represents the expectation value, and $f(x)$ is the underlying probability density function. For the KDE curves shown in the bottom panels of Figures \ref{fig:alphak_mintctff_all} and \ref{fig:alphak_mintctff_restricted}, we used the KDE tools from the  \texttt{scikit-learn} Python package \citep{scikit-learn}.

Figure \ref{fig:alphak_mintctff_all} shows results for the HQ subsample.  The general trends for multiphase gas extent align with the model predictions, but the distributions of $\alpha_K$ are broad, especially in the NE and N categories.  We find that galaxies with extended multiphase gas (E) tend to have entropy profile slopes close to $\alpha_K = 2/3$, as predicted, and single phase galaxies with no extended multiphase gas (N) tend to have greater values of $\alpha_K$.  However, the distribution function for the N galaxies significantly overlaps with the one for the E galaxies and extends below $\alpha_K = 0.5$. In the $\min (t_{\rm cool}/t_{\rm ff})$ dimension, we find that the E galaxies generally have the lowest values, mostly in the $5 \lesssim \min(t_{\rm cool}/t_{\rm ff}) \lesssim 20$ range.  The N galaxies generally have the greatest values, mostly with $\min(t_{\rm cool}/t_{\rm ff}) > 20$, while the NE galaxies are intermediate between the other two categories. All of those findings are similar to those of \citet{Voit2015b}, but in a sample almost four times larger.

Two vertical lines in the bottom panel of Figure \ref{fig:alphak_mintctff_all} show mean values of $\alpha_K$ predicted for the E and N galaxies by the simple analytical model.  They were obtained by plugging the average value of $\sigma_v$ for each galaxy population into equation (\ref{eq:alpha_sigma}).  In both cases, the predicted value is close to the peak of the corresponding KDE curve. 

Figure \ref{fig:alphak_mintctff_restricted} shows results for the low $K_0$ subset. Note that removing the galaxies with elevated central entropy, which complicates measurements of $\alpha_K$, has eliminated the set of single phase (N) galaxies with $\alpha_K \lesssim 2/3$.  All of the remaining single phase galaxies have $\alpha_K \geq 0.8$, and they cluster near $\alpha_K \approx 1$.  Likewise, many of the NE galaxies with low $\alpha_K$ have been pruned from the sample, shifting their mean value of $\alpha_K$ to well above 2/3. Obtaining cleaner measurements of $\alpha_K$ by removing galaxies with $K_0 > 3 \, {\rm keV \, cm^2}$ has therefore sharpened the distribution functions in $\alpha_K$. The galaxies with extended multiphase gas are now clearly distinct from the other two sets in the $\alpha_K$ dimension and cluster near $\alpha_K \approx 2/3$, in agreement with model predictions. 

Two solid vertical lines in the bottom panel of Figure \ref{fig:alphak_mintctff_restricted} show predictions for $\alpha_K$ determined as in Figure \ref{fig:alphak_mintctff_all} for the E and N galaxies by the simple analytical model.  The red vertical line representing the prediction for the N galaxies is quite close to the peak of the corresponding KDE curve, but the blue vertical line representing the prediction for the E galaxies is now near $\alpha_K \approx 0.8$.  This displacement of the prediction away from the peak of the KDE curve (near $\alpha_K \approx 2/3$) for the E galaxies is due entirely to M87, in which $\sigma_v$ is much greater than in other four galaxies. Removing it from that set results in a prediction represented by the dotted blue line, which is much closer to the peak of the KDE curve, indicating good agreement with the analytical model. 

\newpage

\subsection{Comments on Individual Galaxies}
\label{comments_on_indiv_galaxies}

The comparisons shown in Figures \ref{fig:sigma_v_alpha_all_galaxies} through \ref{fig:alphak_mintctff_restricted} are encouraging for the black hole feedback valve model, but there are also some distinct outliers that disagree with the model's predictions.  The presence of outliers should not be surprising given the simplicity of the analytic model.  We do not necessarily expect it to describe all galaxies, especially those in which something clearly more complex is happening. This section  discusses some of the nonconforming galaxies among the low $K_0$ subset and considers characteristics of their environments that may explain why they do not conform.

\subsubsection{M87}\label{M87}

M87 (listed as NGC 4486 in Table \ref{tab:galaxy_params}) has $\sigma_v = 323.0 \pm 4.3$, $\alpha_K$(1--10~kpc)$ = 0.61 \pm 0.1$, and extended multiphase gas.  Its low value of $\alpha_K$, given the high $\sigma_v$, therefore makes it an outlier from the analytic model.  However, its status as the central galaxy in a large potential well (the Virgo galaxy cluster) makes it unusual among the galaxies in the HQ subsample.  The surrounding intracluster medium exerts a larger pressure on this galaxy's atmosphere, causing radiative cooling to exceed SN~Ia heating within the galaxy, thereby violating the assumptions on which equation (\ref{eq:alpha_sigma}) is based.  Instead, we expect this galaxy's atmosphere to be in a precipitation-limited state, with $\alpha_K \approx 2/3$, extended multiphase gas, and $5 \lesssim \min(t_{\rm cool}/t_{\rm ff}) \lesssim 20$.  Its observed characteristics generally agree with those expectations, yet its $\min(t_{\rm cool}/t_{\rm ff}) = 22.73 \pm 0.27$ is slightly greater than expected for a galaxy with extended multiphase gas. 

\subsubsection{NGC 4636} \label{sec:N4636}

NGC 4636 has $\sigma_v = 199.5 \pm 2.7$, $\alpha_K$(1--10~kpc)$ = 1.00 \pm 0.02$, $\min(t_{\rm cool}/t_{\rm ff}) = 10.79 \pm 0.36$, and multiphase gas that does not extend beyond 2~kpc. It is an outlier because its entropy slope is unusually large among the galaxies with $\sigma_v < 240 \, {\rm km \, s^{-1}}$.  \citet{Voit2015b} showed that this galaxy's atmosphere is plausibly precipitation-limited, and \citet{Voit2019} pointed out that its inner entropy profile slope is consistent with that of a pure cooling flow (i.e. $\alpha_K = 1$).  At $\sim 0.1$~kpc, its entropy level reaches $\sim 1 ~\mathrm{keV~cm}^2$, considerably below the level expected from $\sim10^{42} \, \mathrm{erg~s}^{-1}$ of intermittent kinetic feedback power (see \citealt{Voit2019} for details). 

There are several possible explanations for this low central entropy level: (1) time-averaged kinetic AGN power has been $\sim10^{41} \, \mathrm{erg~s}^{-1}$ for the last $\sim100$ Myr, (2) the AGN power has been highly collimated, as in NGC 4261, and has penetrated to $\gg 1$ kpc without dissipating much power, or (3) AGN power has been too weak to balance cooling for the last $\sim100$ Myr. In this last case, a cooling catastrophe may be imminent, as suggested by the entropy profile between 0.5 and 2 kpc, and will soon trigger a strong feedback episode.

\subsubsection{NGC 1521}

NGC 1521 has $\sigma_v = 233.6 \pm 8.9$, $\alpha_K$(1--10~kpc)$ = 0.38 \pm 0.39$, $\min(t_{\rm cool}/t_{\rm ff}) = 20.90 \pm 0.51$, and multiphase gas that does not extend beyond 2~kpc.  The best-fitting value of $\alpha_K$ is lower than the model prediction, but its uncertainty is large, most likely because of low spatial resolution. Its entropy profile has only four radial bins between 1 and 10 kpc, and one additional radial bin interior to 1 kpc. The resulting uncertainty does overlap the analytic prediction. Therefore, improved spatial resolution is necessary to determine if the galaxy does indeed conform to the model.

\subsubsection{NGC 4125}

NGC 4125 has $\sigma_v = 239.8 \pm 6.9$, $\alpha_K$(1--10~kpc)$ = 0.13 \pm 0.45$, and $\min(t_{\rm cool}/t_{\rm ff}) = 28.22 \pm 12.35$. Its multiphase gas classification is unknown.  Like NGC 1521 its apparently shallow entropy slope, relative to the model prediction, may also result from poor spatial resolution. However, the shape of its entropy profile and some of its other characteristics are a bit more interesting. Its X-ray luminosity (measured inside 10 kpc) is the lowest in the HQ sample ($0.023\pm0.001\times10^{42}~\mathrm{erg~s^{-1}}$). Between 1 and 10 kpc, the entropy profile is almost flat, based on six radial bins. However, interior to 1 kpc, the entropy profile slope is much steeper, and \citet{Lakhchaura2018} find a power-law component in the X-ray spectrum, indicating the presence of an AGN with luminosity $0.006\pm0.001\times 10^{41} \mathrm{erg~s^{-1}}$. \citet{Wiklind1995} placed an upper limit on the molecular gas content, but the measurement is uncertain due to high systematic errors. The combination of the presence of an AGN, $\sigma_v~<~240~\mathrm{km~s^{-1}}$, and the flattened entropy profile at larger radii indicate that this may be a galaxy in which the steady flow is cooling-dominated at larger radii and prone to developing entropy inversions \citep{Voit2019}.

\subsubsection{NGC 1404}

NGC 1404 has $\sigma_v = 229.7 \pm 3.8$, $\alpha_K$(1--10~kpc)$ = 0.80 \pm 0.03$, $\min(t_{\rm cool}/t_{\rm ff}) = 20.23 \pm 0.51$, and no extended multiphase gas.  It is notable as the only galaxy in the low $K_0$ subset with $\sigma_v < 240 \, {\rm km \, s^{-1}}$ and no multiphase gas.  Aside from NGC 4636, it has the steepest entropy profile slope among the galaxies with $\sigma_v < 240 \, {\rm km \, s^{-1}}$, which may account for its lack of extended multiphase gas. Furthermore, its entropy profile sharply steepens beyond 7 kpc. It is a satellite of NGC 1399, and so the steep entropy gradient could potentially result from ram pressure stripping that truncates the low-entropy outflow emanating from NGC 1404 as it orbits through the higher entropy intragroup medium around NGC 1399. 

\subsubsection{NGC 533}

NGC 533 has $\sigma_v = 271.9 \pm 5.6$, $\alpha_K$(1--10~kpc)$ = 0.88 \pm 0.14$, $\min(t_{\rm cool}/t_{\rm ff}) = 12.28 \pm 3.75$, and extended multiphase gas.  Along with M87, it is one of only two galaxies in the HQ sample with $\sigma_v > 245 \, {\rm km \, s^{-1}}$ that \citet{Lakhchaura2018} classified as E.  But unlike M87, its entropy slope is completely consistent with the prediction of the simple analytical model.  One reason for this blend of features may be that it is the central galaxy of a larger halo, like M87. According to \citet{ZabludoffMulchaey1998}, the velocity dispersion of the galaxies surrounding NGC 533 is $\sim$464 $\mathrm{km~s^{-1}}$.  Furthermore, its low value of $\min(t_{\rm cool}/t_{\rm ff})$ suggests that the atmosphere of NGC~533 is susceptible to precipitation.  It is also possible that NCG 533 is more similar to galaxies in the NE class, as its status as a galaxy with line emission extending beyond 2~kpc is marginal.  The VIMOS map of that galaxy presented in \citet{Hamer_2016MNRAS.460.1758H} shows just a single small H$\alpha$ blob lying not far outside the 2~kpc radius limit that distinguishes NE galaxies from E galaxies.

\subsection{Predictions for Profile Normalization}\label{P_ne_eq}

\begin{figure*}
    \centering
    \includegraphics[width=\textwidth]{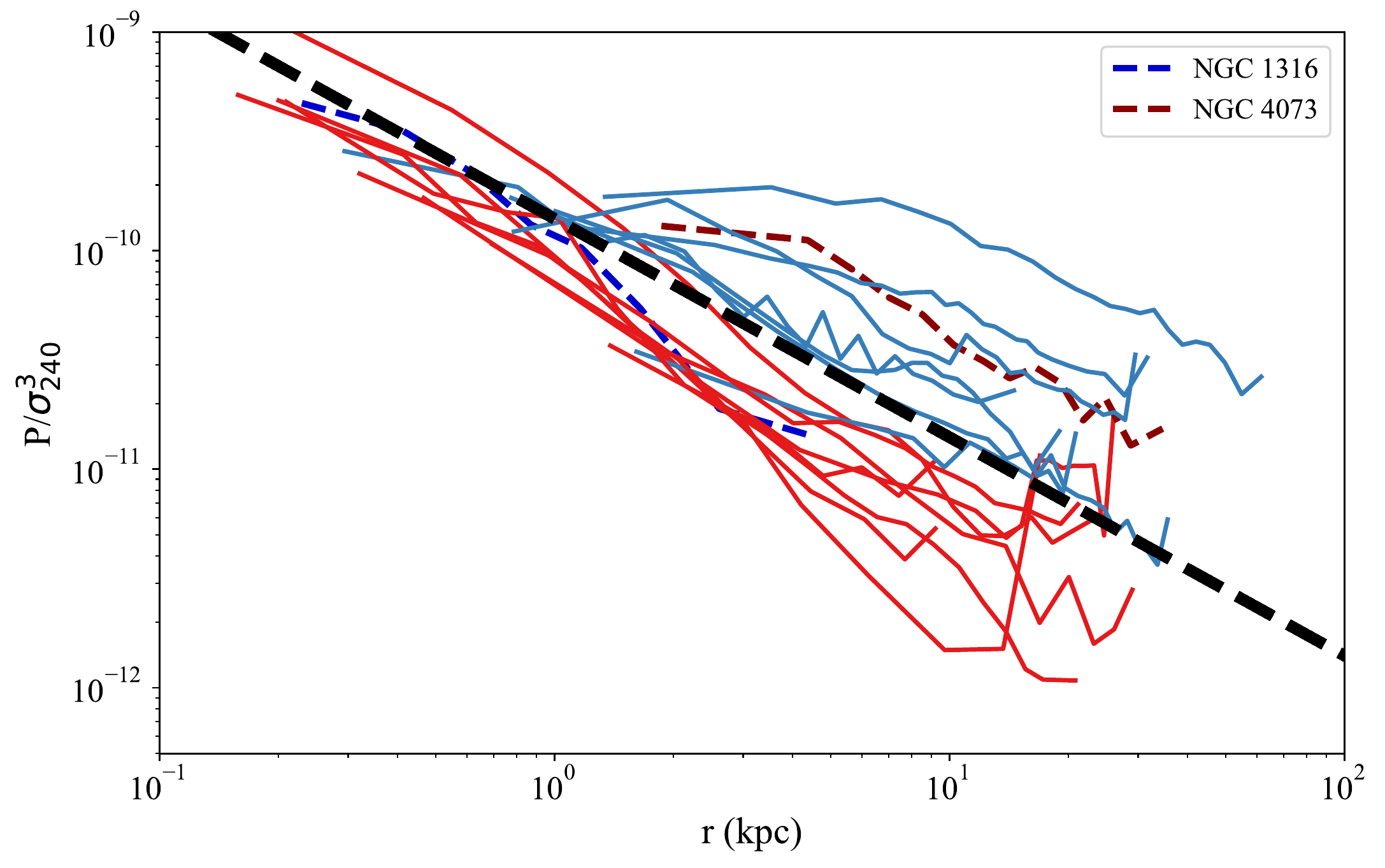}
    \caption{\textbf{Scaled pressure profile comparisons for single phase and multiphase galaxies in the HQ subsample.} A thick dashed black line shows the scaled pressure profile ($P/\sigma_{240}^3$) from equation (\ref{eq:Peq_scaled}), along which SN~Ia heating should approximately balance radiative cooling.  According to the model of \citet{Voit2019}, galaxies with confining pressures greater than the line at $\sim 10$~kpc should be prone to precipitation and formation of extended multiphase gas, while galaxies with confining pressures below the line should be free of extended multiphase gas because of SN~Ia-heated outflows that focus cooling and condensation within the central $\sim 1$~kpc.  The observed pressure profiles of extended multiphase (E, blue lines) and single-phase (N, red lines) galaxies, shown without error bars for clarity, generally conform to that pattern. There are two exceptions in the HQ subsample: NGC 1316 (dark blue dashed line) and NGC 4073 (dark red dashed line). See Section \ref{P_ne_eq} for discussions of these two galaxies.\label{fig:Peq_r} }
\end{figure*}

\begin{figure*}
    \centering
    \includegraphics[width=\textwidth]{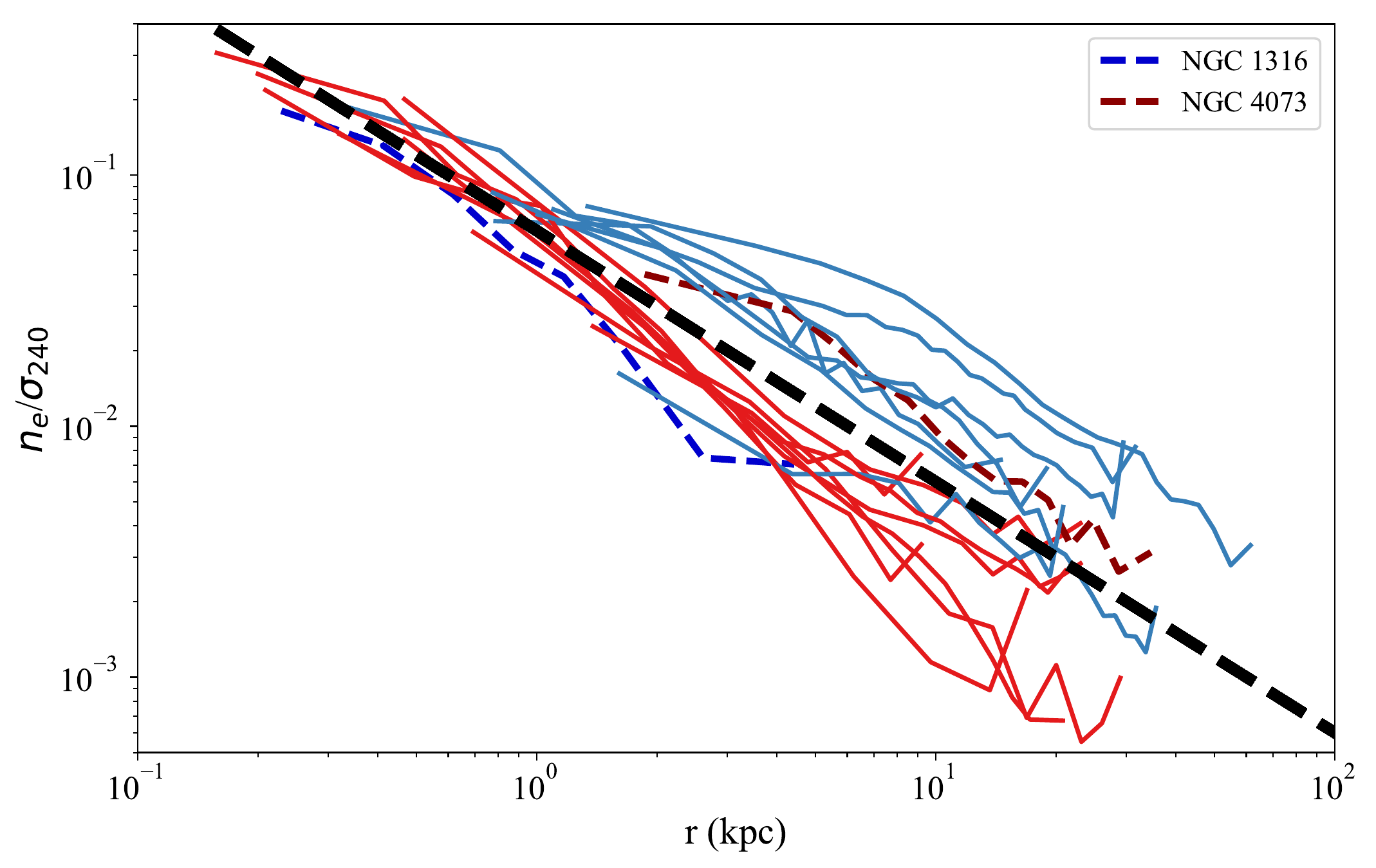}
    \caption{\textbf{Scaled electron density comparisons for single phase and multiphase galaxies in the HQ subsample.} A thick dashed black line shows the scaled electron density profile ($n_e/\sigma_{240}$) from equation (\ref{eq:neleq_scaled}), along which SN~Ia heating should approximately balance radiative cooling.  All other figure elements are as in Figure \ref{fig:Peq_r} and follow the same basic pattern, with low density allowing SN~Ia heating to exceed radiative cooling in the 1--10~kpc range in single phase galaxies, while high density prevents SN~Ia heating from exceeding radiative cooling in multiphase galaxies. NGC 1316 and NGC 4073 remain exceptions (see Section \ref{P_ne_eq}).}\label{fig:ne_eq_r}
\end{figure*}

The analytical prediction from \citet{Voit2019} expressed in equation (\ref{eq:alpha_sigma}) presumes that SN~Ia heating exceeds radiative cooling in the 1--10~kpc region.  Figure \ref{fig:sigma_v_alpha_restricted} shows that the $\sigma_v$--$\alpha_K$ relationship observed in the low $K_0$ subset supports that prediction, especially in the interval $220 \, {\rm km \, s^{-1}} \leq \sigma_v \leq 300 \, {\rm km \, s^{-1}}$, and suggests another test of the model.  For equation (\ref{eq:alpha_sigma}) to be a correct explanation for the $\sigma_v$--$\alpha_K$ relation, the normalizations of the pressure and density profiles in single phase galaxies must be consistent with an excess of SN~Ia heating over radiative cooling at $\sim$1--10~kpc.  This section assesses that prediction by comparing SN~Ia heating with radiative cooling in the low $K_0$ subset using observed density and pressure profiles from \citet{Lakhchaura2018}, focusing on the N and E galaxies.

According to the model, there is a critical gas pressure profile along which radiative cooling equals SN~Ia heating:
\begin{equation}
    \label{eq:Peq}
    P_{\mathrm{eq}}(r)\equiv\Big[\Big(\epsilon_*+\frac{3}{2}\sigma_v^2\Big)\Big(\frac{n^2}{n_en_p}\Big)\frac{\rho_*}{t_*\Lambda(T)}\Big]^{1/2}kT
    \; \; .
\end{equation}
This is Equation (11) from \citealt{Voit2019}, in which $n_p$ is the proton density, $\rho_*$ is the stellar mass density, $t_*^{-1}$ is the specific stellar mass-loss rate, and $\Lambda(T)$ is the radiative cooling function. For the velocity dispersion and temperature corresponding to the critical entropy profile slope ($\alpha_K = 2/3$ at $\sigma_v\approx240~\mathrm{km~s^{-1}}$), the model predicts $kT\approx0.75~\mathrm{keV}$ and the critical pressure and density profiles for solar metallicity gas become:
\begin{eqnarray}
    P_{\mathrm{eq}}(r) & \: \approx \: & (1.4\times10^{-10}\mathrm{erg ~cm^{-3}})~\sigma_{240}^3r_{\mathrm{kpc}}^{-1} 
        \label{eq:Peq_scaled}
\\
    n_{e,\mathrm{eq}} & \: \approx \: & (0.06~\mathrm{cm^{-3}})~\sigma_{240}r_{\mathrm{kpc}}^{-1},
        \label{eq:neleq_scaled}
\end{eqnarray}
where $r_{\mathrm{kpc}}$ is radius in kiloparsecs, if the weak dependence of $\Lambda(T)$ on $\sigma_v$ is ignored (see equations 12 and 13 in \citealt{Voit2019}). Those expressions assume an isothermal stellar mass distribution ($\rho_*=\sigma_v/2\pi Gr^2$) and the fiducial values $\mu m_p\epsilon_*\approx 2~\mathrm{keV}$ and $t_*\approx 200~\mathrm{Gyr}$ (see \citealt{Voit2019} for explanations of those fiducial values).

Figures \ref{fig:Peq_r} and \ref{fig:ne_eq_r} show comparisons of the extended multiphase (E) and single phase (N) galaxies in the HQ sample with the equilibrium pressure and density profiles from the model. Galaxies with multiphase gas confined to the central 2 kpc (NE) have been removed for clarity. If the model applies to the galaxies in the figure, then the equilibrium profiles derived from the model should divide the observed profiles of galaxies with extended multiphase gas (which typically have $\alpha_K \approx 2/3$) from those of galaxies with no extended multiphase gas (which typically have $\alpha_K \approx 1$). The black hole feedback valve model predicts that single phase galaxies (N) should have pressure profiles below $P_{\rm eq}$ at $\sim$1--10~kpc, while multiphase galaxies (E) should have pressure profiles above $P_{\rm eq}$ in that radial range.

The figures show that the equilibrium profiles ($P_{\mathrm{eq}}$ and $n_{e,\mathrm{eq}}$) do indeed divide the HQ sample as predicted. However, there are two notable exceptions, one each among the multiphase and single phase galaxies: NGC 1316 and NGC 4073. 

The multiphase galaxy NGC 1316 has a pressure profile that is close to $P_{\rm eq}$ within $\sim 1$~kpc but that drops below the $P_{\rm eq}$ locus at $\sim$1--4~kpc. It therefore has unusually low circumgalactic pressure for an early-type galaxy with extended multiphase gas. That said, it has apparently experienced a large feedback event during the last 0.5~Gyr \citep{Lanz2010}. Its radio and X-ray properties indicate that a kinetic feedback outburst of $\sim 5 \times 10^{58} \, {\rm ergs}$ occurred $\sim 0.4$~Gyr ago, producing large buoyant cavities in the galaxy's hot atmosphere. Those cavities are now $\sim 100$~kpc from the center and may have stimulated production of extended multiphase gas through uplift \citep[e.g.,][]{McNamara2016}.

Conversely, the single phase galaxy NGC 4073 has one of the greatest X-ray luminosities in the HQ subsample, along with one of the hottest temperatures, and lies above the $P_{\rm eq}$ locus. It has $\alpha_K\approx0.6$ and $\sigma_v\approx268~\mathrm{km~s^{-1}}$ and does not conform to the analytic model within the measurement uncertainty. However, it also has $\min (t_{\rm cool} / t_{\rm ff}) = 32.22 \pm 2.92$, which is large enough to strongly suppress multiphase condensation and would be an unusually large value of $\min (t_{\rm cool} / t_{\rm ff}$) for a galaxy with extended multiphase gas. 

\section{Conclusions}\label{conclusions}

This paper has presented evidence for a relationship between stellar velocity dispersion $\sigma_v$ and entropy profile slope $\alpha_K$ at 1--10~kpc in the \citet{Lakhchaura2018} sample of early-type galaxies.  The characteristics of that relationship align with the predictions of the analytic model proposed by \citet{Voit2019}, which has no free parameters. Furthermore, the significance of the result and its agreement with the analytic model improve as restrictions are placed on the sample to exclude galaxies that have properties inconsistent with the analytical model's assumptions.

In contrast to previous studies of this type, we applied stricter limits on the data quality of the archival observations as well as limits on the characteristics of the galaxies being considered.  Our HQ subsample of the  \citet{Lakhchaura2018}, which contains 36 galaxies, requires their radial entropy profiles to have at least three entropy bins in the 1--10 kpc range, outside of which the analytical model is not expected to apply.  We measured the entropy profiles in that radial range by fitting the three-parameter model $K(r) = K_0 + K_{10} (r / 10 \, {\rm kpc})^{\alpha_K}$.  After obtaining those fits, we characterized the resulting $\sigma_v$--$\alpha_K$ relationship by fitting it with a two-parameter linear model. For the HQ subsample, we found $\alpha_K (240 \, {\rm km \, s^{-1}}) = 0.70 \pm 0.15$, in excellent agreement with the analytical prediction of $\alpha_K (240 \, {\rm km \, s^{-1}}) \approx 2/3$, along with evidence for a rise in $\alpha_K$ with $\sigma_v$ at the $\approx 2 \sigma$ level.  

However, more than a third of the galaxies in the HQ subsample have $K_0 > 3 \, {\rm keV \, cm^2}$, a feature that compromises the fits for $\alpha_K$ and indicates that the 1--10~kpc region is heated by an AGN in addition to the SN~Ia heating assumed by the analytical model.  We therefore selected a low $K_0$ subset by excluding galaxies with $K_0 > 3 \, {\rm keV \, cm^2}$.  Fitting a linear $\sigma_v$--$\alpha_K$ relationship to the remaining 22 galaxies gives $\alpha_K (240 \, {\rm km \, s^{-1}}) = 0.74 \pm 0.27$, consistent with with the analytical prediction, and provides stronger evidence for a rise in $\alpha_K$ with $\sigma_v$ at the $2.4 \sigma$ level.

The final cut we made was to restrict the range of $\sigma_v$ to galaxies for which the analytic model is most likely to be accurate.  We excluded the one galaxy with $\sigma_v < 220 \, {\rm km \, s^{-1}}$, because the analytical model is not self-consistent there.  We also excluded the five galaxies that have $K_0 < 3 \, {\rm keV cm^2}$ and $\sigma_v > 300 \, {\rm km \, s^{-1}}$ because the analytic predictions of \citet{Voit2019} overshoot their numerical predictions in that range of $\sigma_v$. For the remaining 16 galaxies we found $\alpha_K (240 \, {\rm km \, s^{-1}}) = 0.66 \pm 0.19$ and more conclusive evidence for a rise in $\alpha_K$ with $\sigma_v$ at the $3.5 \sigma$ level.  Furthermore, the best-fitting linear $\sigma_v$--$\alpha_K$ relationship is nearly identical to the analytic prediction.

The analytic model also successfully predicts how the normalizations of pressure and density profiles observed among galaxies with extended multiphase gas differ from the profiles observed among galaxies without extended multiphase gas.  According to the model, galaxies with low circumgalactic pressure and high $\sigma_v$ should be free of extended multiphase gas, while galaxies with high circumgalactic pressure should be precipitation-limited and prone to developing multiphase gas.  The galaxies in our HQ subsample conform to that prediction with two exceptions: NGC 1316 and NGC 4073 (see Figures \ref{fig:Peq_r} and \ref{fig:ne_eq_r}). NGC 1316 has an unusually low pressure at $< 10$~kpc for a multiphase galaxy but contains a powerful radio source (Fornax A).  NGC 4073 has an unusually high pressure atmosphere for a galaxy without extended multiphase gas but also has $t_{\rm cool} / t_{\rm ff} > 30$, which may account for the absence of a cooler gas phase.

Our work shows that while the \citet{Voit2019} analytic model may be overly idealized, it describes the relationship between key galaxy parameters well and can be used to further our understanding of how feedback in massive galaxies works. The comparison of the model to the data supports the notion that SN~Ia heating plays an important role in the thermal evolution of massive galaxies. Furthermore, the relationship between entropy profile slope and velocity dispersion is highly dependent on the external gas pressure at larger radii. Current X-ray observations are not able to resolve pressure measurements at large radii, but Athena and Lynx may be able to. Given the model predictions and existing observations, one could predict what the gas pressure at large radii should be and then test that prediction with the next generation of X-ray telescopes.

\begin{acknowledgements}
The scientific results reported in this article are based on data obtained from the Chandra Data Archive. This research has made use of software provided by the Chandra X-ray Center in the applications package CIAO \citep{Fruscione_CIAO_2006SPIE.6270E..1VF}. N.W. is supported by the GACR grant 21-13491X.
\end{acknowledgements}

\end{document}